%% file: OosterDichot.tex
\documentclass[twocolumn]{aastex63}
\input{defs.tex}

\accepted{July 1, 2021}

\submitjournal{ApJ}

\shorttitle{RR Lyrae as Galactic probes: IV. New insights into and around the Oosterhoff dichotomy.}
\shortauthors{M. Fabrizio et al.}

\begin{document}
\title{On the use of field RR Lyrae as Galactic probes: IV. \\ 
New insights into and around the Oosterhoff dichotomy.
\footnote{Based on observations obtained with the du Pont telescope at Las
Campanas Observatory, operated by Carnegie Institution for Science. Based in
part on data collected at Subaru Telescope, which is operated by the National
Astronomical Observatory of Japan. Based partly on data obtained with the STELLA
robotic telescopes in Tenerife, an AIP facility jointly operated by AIP and IAC.
Some of the observations reported in this paper were obtained with the Southern
African Large Telescope (SALT). Based on observations made with the Italian
Telescopio Nazionale Galileo (TNG) operated on the island of La Palma by the
Fundaci\'on Galileo Galilei of the INAF (Istituto Nazionale di Astrofisica) at
the Spanish Observatorio del Roque de los Muchachos of the Instituto de
Astrofisica de Canarias. Based on observations collected at the European
Organisation for Astronomical Research in the Southern Hemisphere.}}

\correspondingauthor{Michele Fabrizio}
\email{michele.fabrizio@ssdc.asi.it}

\author[0000-0001-5829-111X]{M.~Fabrizio}
\affil{INAF - Osservatorio Astronomico di Roma, Via Frascati 33, 00078, Monte Porzio Catone (Roma), Italy}
\affil{Space Science Data Center - ASI, Via del Politecnico s.n.c., 00133 Roma, Italy}
\input{authors.tex}

\begin{abstract}

We discuss the largest and most homogeneous spectroscopic dataset of field RR
Lyrae variables (RRLs) available to date. We estimated abundances using both
high-resolution and low-resolution (\deltaS\ method) spectra for fundamental
(RRab) and first overtone (RRc) RRLs. The iron abundances for 7,941 RRLs were
supplemented with similar literature estimates available, ending up with 9,015
RRLs (6,150 RRab, 2,865 RRc). The metallicity distribution shows a mean value of
$\langle\feh\rangle=-1.51\pm0.01$, and $\sigma$(standard deviation)$=0.41$~dex
with a long metal-poor tail approaching $\feh\simeq-3$ and a sharp metal-rich
tail approaching solar iron abundance. The RRab variables are more metal-rich
($\langle\feh\rangle_{ab}=-1.48\pm0.01$, $\sigma=0.41$~dex) than RRc variables
($\langle\feh\rangle_{c}=-1.58\pm0.01$, $\sigma=0.40$~dex).
The relative fraction of RRab variables in the Bailey diagram (visual amplitude
\textit{vs} period) located along the short-period (more metal-rich) and the
long-period (more metal-poor) sequences are 80\%\ and 20\%, while RRc variables
display an opposite trend, namely 30\%\ and 70\%.
We found that the pulsation period of both RRab and RRc variables steadily
decreases when moving from the metal-poor to the metal-rich regime. The visual
amplitude shows the same trend, but RRc amplitudes are almost two times more
sensitive than RRab amplitudes to metallicity.
We also investigated the dependence of the population ratio (N$_c$/N$_{tot}$) of
field RRLs on the metallicity and we found that the distribution is more complex
than in globular clusters. The population ratio steadily increases from
$\sim$0.25 to $\sim$0.36 in the metal-poor regime, it decreases from $\sim$0.36
to $\sim$0.18 for $-1.8\le\feh\le-0.9$ and it increases to a value of $\sim$0.3
approaching solar iron abundance.

\end{abstract}


\section{Introduction} \label{sec:intro}

The Oosterhoff dichotomy is one of the most interesting problems in modern
astrophysics. It was first described by the seminal work of
\citet{oosterhoff39}, where he investigated the period distribution of cluster
RR Lyrae (RRLs) and found that Galactic globular clusters (GCs) hosting RRLs can
be split into two different groups. The so-called Oosterhoff type II (OoII)
clusters have a mean period for fundamental mode RRLs (RRab) of $\langle
P_{ab}\rangle\simeq0.651$~days and a mean period for first overtone RRLs (RRc)
of $\langle P_c\rangle\simeq0.356$~days, while the Oosterhoff type I (OoI)
clusters have a mean period for RRab of $\langle P_{ab}\rangle\simeq0.557$~days
and a mean period for RRc of $\langle P_c\rangle\simeq0.312$~days
\citep{vanAgt59,cacciari76,sandage81b,lee90,bono16}. This finding was further
strengthened by the spectroscopic evidence that OoI GCs are more metal-rich and
cover a broad range in metal abundances, while OoII GCs are more metal-poor
stellar systems \citep{arp55,sandage60}. These empirical separations show up
very clearly in the so-called Bailey diagram (luminosity amplitude \textit{vs}
logarithmic period), indeed, OoII GCs attain, at fixed luminosity amplitude,
pulsation periods that are systematically longer than OoI GCs.

\begin{figure} 
\centering
\includegraphics[width=1\columnwidth]{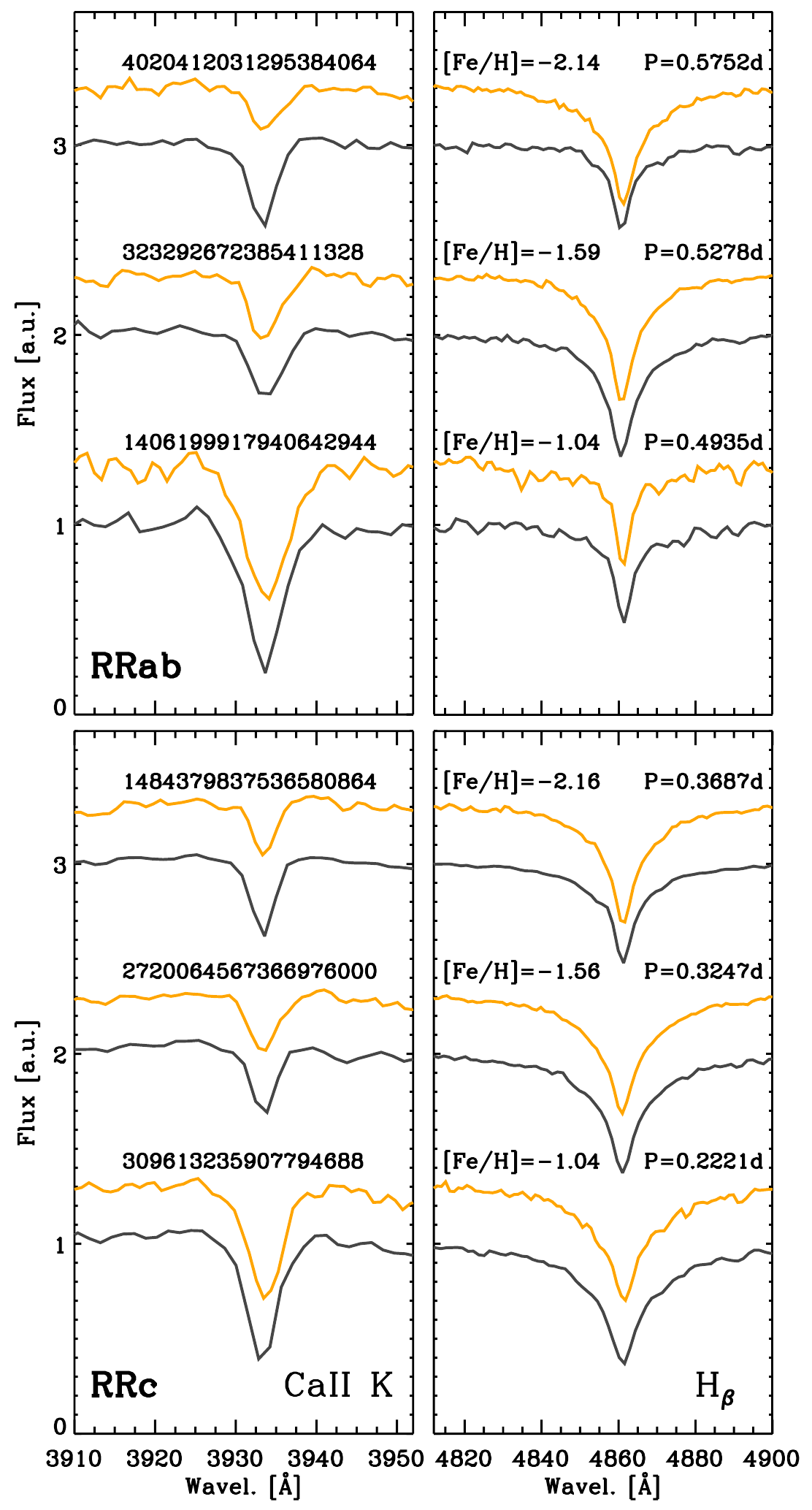} 
\caption{\textit{Top:} selected low-resolution spectra for three fundamental
RRLs with different iron abundances (see labeled values) from the LAMOST
(orange) and the SEGUE (grey) datasets. The $Gaia$~EDR3~IDs are also labelled.
The left panels display the region across the \caiik\ line, while the right ones
the region across the \hbeta\ line. \textit{Bottom:} Same as the top, but for
three RRc variables.}\label{fig:spectra}
\end{figure} 

Subsequent investigations brought forward that the RRL population ratio, i.e.
the ratio between the number of RRc (N$_c$) variables and the total number of
RRLs (N$_{tot}$ = N$_{ab}$ + N$_c$)\footnote{Note that in the following we are
considering mixed mode variables (RRd) together with first overtone variables,
because the dominat mode is typically the first overtone.} is, together with
mean period, the most popular pulsation diagnostic to dictate the difference
between OoI and OoII GCs \citep{stobie71,castellani87}. Indeed, the population
ratio for OoII GCs is N$_c$/N$_{tot}\sim0.44$, while for OoI GCs is
N$_c$/N$_{tot}\sim0.29$ \citep{braga16}. The values of both mean periods and
population ratios typical of OoI and OoII GCs depend on the criteria adopted to
select cluster variables \citep{fiorentino15}, but the quoted estimates are only
marginally affected.

The literature concerning the Oosterhoff dichotomy is vast and includes
theoretical \citep{lee94,bono95,cassisi04}, photometric \citep{lee99} and
spectroscopic \citep{vandenbergh93b} investigations. However, a comprehensive
empirical scenario concerning the Oosterhoff dichotomy in field and cluster RRLs
was built over half century by \citet[][and references therein]{sandage10}. This
is the reason why the same problem is known in the literature as the
"Oosterhoff-Arp-Sandage" period-shift effect \citep[][and references
therein]{catelan15}.

During the last few years the large photometric survey of nearby dwarf galaxies
breathed new life into this classical problem. RRLs in Local Group galaxies and
in their GCs have mean fundamental periods that fill the so-called Oosterhoff
gap, indeed they attain intermediate values ($0.58<P_{ab}<0.62$~days) between
OoI and OoII GCs \citep{petroni03,catelan09}. This circumstantial evidence
indicates that the environment affects the Oosterhoff dichotomy
\citep{fiorentino15}. Moreover and even more importantly, detailed and
comprehensive investigations of three metal-rich clusters (NGC~6388,
\citealt{pritzl02}; NGC~6441, \citealt{pritzl03}; NGC~6569, \citealt{baker07})
and of the most massive Galactic GC ($\omega$~Cen, \citealt{braga16}) further
strengthened the possible occurrence of additional Oosterhoff groups.

A new spin concerning this classical problem was recently provided by
\citet{fabrizio19}. They provided new metallicity estimates using low resolution
spectra collected by SEGUE \citep{lee08} for more than 3,000 field RRLs. This
means an increase by almost one order of magnitude with similar data available
in the literature. They found that the Oosterhoff dichotomy was mainly caused by
the fact that metal-intermediate GCs lack of sizeable samples of RRLs. Indeed,
field RRLs display a steady variation in the period distribution and in the
Bailey diagram when moving from the metal-poor to the metal-rich regime.
However, this investigation was hampered by two limitations: \textit{i)} the
analysis was only based on fundamental variables; \textit{ii)} they did not
investigate the dependence of the population ratio on the iron content.

In the following we address these key issues by using new homogeneous
metallicity estimates (see Sect.~\ref{sec:measure}) based on high-resolution
spectra and on a new calibration of the \deltaS\ method \citep{crestani21}. This
catalog was supplemented with similar metallicity estimates available in the
literature (see Sect.~\ref{sec:catalog}). As a whole we ended up with a
spectroscopic catalog including 9,015 RRLs (6,150 RRab, 2,865 RRc) with at least
one metallicity estimate. In Section~\ref{sec:iron} we discuss the metallicity
distribution function of the spectroscopic catalog and investigate the
difference between RRc and RRab variables. Section~\ref{sec:bailey} deals with
the distribution of the spectroscopic catalog into 2D and 3D realisations of the
Bailey diagram. In this section are also discussed the physical mechanisms affecting
the Bailey diagram, and in particular, the role played by the Blazhko
phenomenon. Section~\ref{sec:metalbins} is focussed on the dependence
of the pulsation period and visual amplitude on metallicity. 
The relation between the population ratio and the metallicity for
field RRLs is analysed in Section~\ref{sec:ratio}, together with a detailed
comparison of cluster RRLs and their horizontal branch morphology.
Section~\ref{sec:summary} includes the summary of the current findings and a few
brief remarks concerning the future developments of this project.

\section{Metallicity measurement}\label{sec:measure}

\begin{table*}
\caption{RR Lyrae spectroscopic datasets. The columns 3, 4, and 5 list the total
number of RRLs, the number of calibrating RRLs and the number of RRLs included
in the spectroscopic catalog. The columns 6 and 7 give the zero-points and the
slopes of the linear relations adopted to transform literature iron abundances
into our metallicity scale. The last column lists the standard deviation of the
fit.}
\label{tbl:rrlcount}
\centering
\begin{tabular}{l r r r r r r r}
\hline
\hline
Source & Priority & N$_{tot.}$ & N$_{cal.}$ & N$_{sel.}$ & $a\pm\epsilon_a$ & $b\pm\epsilon_b$ & $\sigma_{\rm fit}$\\
\hline
      HR our        & 1 &  190 & \ldots & 190 & \ldots          & \ldots          & \ldots \\
      HR lit$^a$    & 2 &   56 & \ldots &  56 & \ldots          & \ldots          & \ldots \\
    \deltaS         & 3 & 7928 & \ldots &7751 & \ldots          & \ldots          & \ldots \\
  \citet{zinn20}   & 4 &  462 & 243    & 219 & $-0.01\pm0.05$ &   $1.03\pm0.03$ & 0.23 \\
   \citet{liu20}   & 5 & 4805 & 4093   & 708 & $-0.41\pm0.10$ &   $0.66\pm0.07$ & 0.23 \\
  \citet{dambis13} & 6 &  399 & 242    &  10 & $-0.01\pm0.05$ &   $1.03\pm0.03$ & 0.25 \\
\citet{duffau14}   & 7 &   59 & 25     &  29 & $ 0.13\pm0.23$ &   $0.90\pm0.13$ & 0.20 \\
  \citet{sesar13}  & 8 &   50 & 24     &  23 & $ 0.14\pm0.33$ &   $0.93\pm0.16$ & 0.12 \\
       RAVE        & 9 &   21 & \ldots &   5 &  \ldots         & \ldots          & \ldots \\
 SEGUE-SSPP        &10 & 2781 & 2756   &  24 & $-0.56\pm0.02$ &   $0.62\pm0.01$ & 0.24 \\
     \hline
\end{tabular}
\tablenotetext{}{$^a$ A compilation from \cite{fernley96}, \cite{lambert96},
\cite{for11}, \cite{liu13}, \cite{nemec13}, \cite{govea14}, \cite{pancino15},
\cite{sneden17}, \cite{chadid17}, \cite{andrievsky18}, as collected and 
normalised by \cite{crestani21}.}
\end{table*}

The metallicity estimates of RRLs were derived by using the most updated version
of the \deltaS\ method. The \deltaS\ method was originally introduced by
\citet{preston59}, and it is based on the comparison of the pseudo-equivalent
width (EW) of the \caiik\ and hydrogen \hbeta, \hgamma, \hdelta\ lines, and it
was used exclusively for the fundamental RRL variables by following the
prescriptions developed by \citet{layden94}. The use of this diagnostic requires
transformation onto a standard EW system by calibrating these measured calcium
and hydrogens EWs with values of a set of spectroscopic "standards" with the
same spectrograph and instrument configuration used to collect the scientific
data. Such an approach has been recently upgraded by \citet{crestani21}
providing a new calibration of the \deltaS\ method, based on a large sample of
high-resolution spectra for more than 140 RRLs. The advantages of this new
calibration are \textit{i)} the extension of the \deltaS\ also to RRc variables,
\textit{ii)} the independence from the transformations between different EW
systems and \textit{iii)} the opportunity to use only one, two or all three
Balmer lines.

\begin{figure*} 
\centering
\includegraphics[width=1.8\columnwidth]{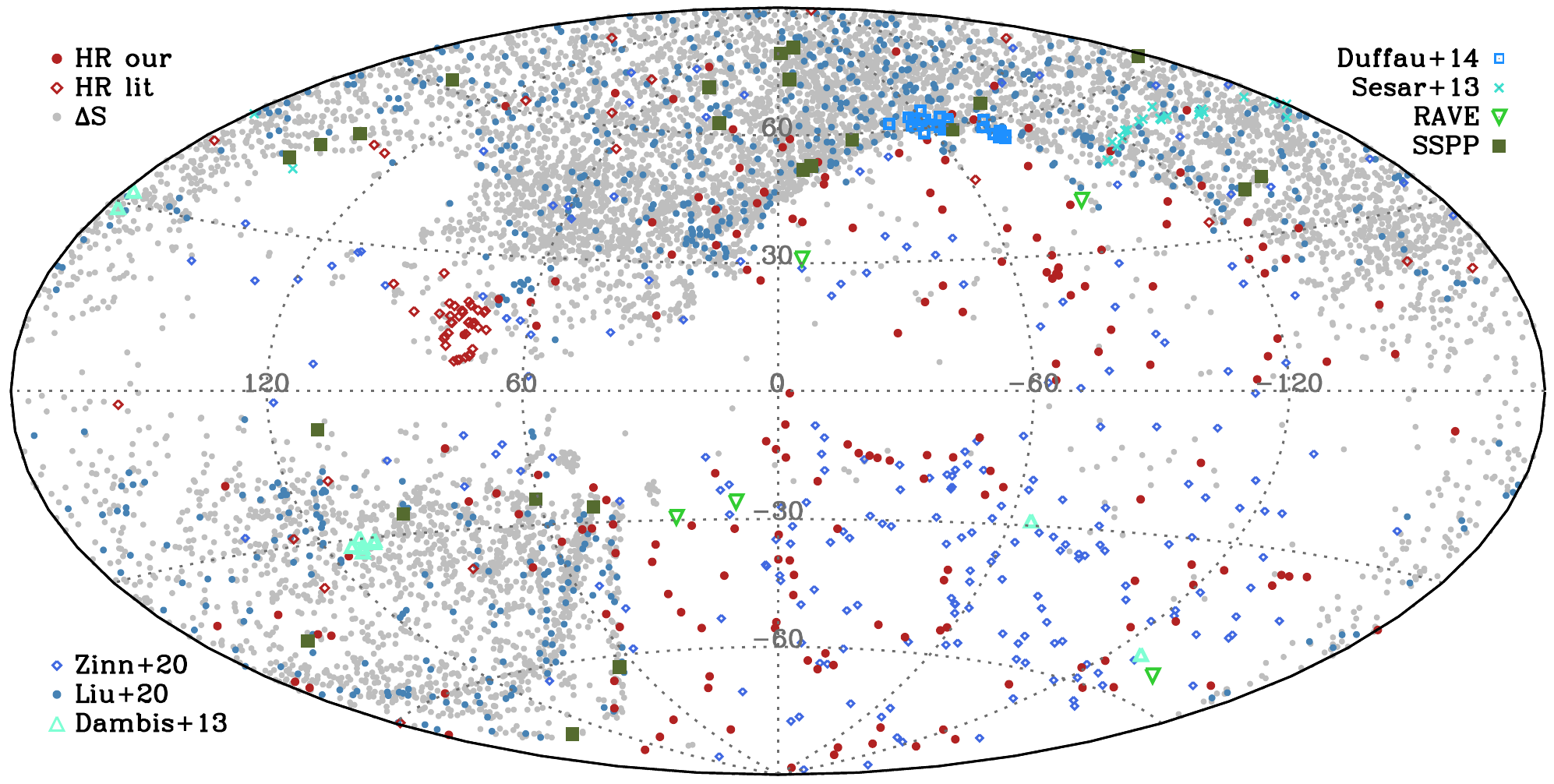} 
\caption{Distribution in Galactic coordinates of the entire spectroscopic sample
(9,015 variables). RRLs coming from different spectroscopic datasets are marked 
with different symbols and/or colours (see labels and Table~\ref{tbl:rrlcount}).}
\label{fig:sky}
\end{figure*} 

In this investigation we used the same spectroscopic sample presented in
\citet{crestani21}. It is mainly based on 5,885 low-resolution ($R\sim2000$)
spectra collected by the Sloan Extension for Galactic Exploration and
Understanding Survey of the Sloan Digital Sky Survey (SEGUE-SDSS,
\citealt{yanny09,alam15})\footnote{\url{https://dr16.sdss.org/optical/spectrum/search}}, 
covering 3,004 RRab and 1,562 RRc stars. 
Moreover, we
took advantage of the huge spectroscopic dataset collected by the Large Sky Area
Multi-Object Fiber Spectroscopic Telescope DR6 (LAMOST, \citealt{cui12,zhao12})
for which data with a resolution of $R\sim2000$ are available. A sample of 5,067
spectra were downloaded from the on-line query
interface\footnote{\url{http://dr6.lamost.org/search}} by using a search radius
of 2.0~arcsec around each target listed in the RRLs photometric catalogue (see
Sect.2 in \citealt{crestani21}), and we ended up with 2,469 RRab and 1,182 RRc
variables with LAMOST spectra (490 in common with SEGUE).
Figure~\ref{fig:spectra} shows representative spectra for the LAMOST (orange)
and the SEGUE (grey) low-resolution (LR) datasets, for three RRab (top panels)
and three RRc (bottom panels), in the region of \caiik\ and \hbeta\ lines, with
different iron abundance. The profiles of \hgamma\ and \hdelta\ are very similar
to that of \hbeta\ and therefore they are not shown. The similarity between
LAMOST and SEGUE datasets allows us to use the same approach and wavelength
limits described in \citet{crestani21} to measure the EWs involved in the
\deltaS\ method by using an updated
\texttt{IDL}\footnote{\url{https://www.harrisgeospatial.com/Software-Technology/
IDL}} version of the
\texttt{EWIMH}\footnote{\url{http://physics.bgsu.edu/~layden/ASTRO/DATA/EXPORT/
EWIMH/ewimh.htm}\label{phd}} program \citep{fabrizio19, layden94}. Additionally, we
similarly applied this method to 178 stars with low signal-to-noise spectra
collected at higher resolution (see below) and degraded to a spectral resolution
of $R\simeq2000$ and sampling $\Delta\log(\lambda)=0.0001$, in order to mimic
the native resolution of the SEGUE data. Finally, we applied the relation
described in Eqn.~1 of \citet{crestani21} to obtain an estimate of
\feh$_{\Delta\rm S}$ for 7,928 RRL variables. 
The reader interested in more a detailed discussion concerning the comparison
between the metallicity scale for field RRLs, based on both high- and
low-resolution spectra, is refereed to Sections 4 and 6 in \citet{crestani21}.

Moreover, we also extended the high-resolution (HR, $R\simeq35,000$) sample ---
including data collected with the echelle spectrographs at du Pont (Las Campanas
Observatory)\footnote{Private communication. They will become available in a few months, because they are associated with a PhD project (Crestani et al. 2021, in preparation).\label{phd}} and at STELLA (Izana Observatory)\textsuperscript{\ref{phd}}, 
UVES and X-shooter at VLT (Cerro Paranal Observatory)\footnote{\url{http://www.eso.org/sci/observing/phase3/data_streams.html}\label{eso}}, 
HARPS at the 3.6m telescope and FEROS at the 2.2m
telescope (La Silla Observatory)\textsuperscript{\ref{eso}}, HARPS-N at the Telescopio Nazionale Galileo
(Roque de Los Muchachos Observatory)\textsuperscript{\ref{phd}}, the HRS at SALT (South African
Astronomical Observatory)\textsuperscript{\ref{phd}}, and the HDS at Subaru (National Astronomical
Observatory of Japan)\footnote{\url{https://stars2.naoj.hawaii.edu/}} --- for 154 RRab and 36 RRc, ending up with 190 RRLs in
the HR sample. The metallicity estimation of these spectra was performed by
following the classical approach as described in \citet{crestani21}.

\section{The RRL spectroscopic catalog}\label{sec:catalog}

As described in Sect.~\ref{sec:measure}, we have estimated the \feh\ of 190 RRLs
by means of HR spectroscopic analysis and of 7,928 RRLs by adopting the \deltaS\
method on LR spectra. In the last 25 years, several papers providing \feh\ for
RRLs were published. Therefore, we have supplemented our own \feh\ estimates
with those from the literature, to build up an extended catalog of spectroscopic
metallicities for RRLs.

To provide a clear picture of the data available in the literature and of the
priority ranking that we are going to adopt, we have separated the \feh\
estimates found in the literature into those coming from either HR or LR
spectroscopy. More specifically, we have collected \feh\ estimates based on HR
data of 56 RRLs from ten different papers (see the references listed in
Tab.~\ref{tbl:rrlcount}) and on LR data of 1,018 RRLs from both RRL-specific
papers and from large spectroscopic surveys like RAVE \citep{steinmetz06} and
SEGUE (from the Stellar Parameter Pipeline - SSPP, \citealt{lee08}).

We ended up with metallicities for 9,015 RRLs. This overall value is smaller
than the sum of the quoted sources for two different reasons. \textit{i)} We
performed a double and in some cases a triple visual check of the spectra for
faint targets ($\G\le 19.5$~mag). We found that a few hundred of them have had
spectra with signal-to-noise ratios that were borderline for a solid metallicity
estimate. They were removed. \textit{ii)} There are a few hundreds of RRLs with
\feh\ estimates from more than one source.
For this reason, we have ranked the priorities of the different 
sources as indicated in Table~\ref{tbl:rrlcount}.

The quoted priority rank is based on the following criteria, sorted by
decreasing relevance: \textit{i)} the highest priority is given to HR
spectroscopic measurements; subsequently to \textit{ii)} our own estimates based
on \deltaS; \textit{iii)} lowest priority to the large datasets. These criteria
were finally weighted by other factors (instrumentation, method adopted,
uncertainties, single \textit{vs} multiple measurements) to provide the final
ranking.

Table~\ref{tbl:rrlcount} provides the number of RRLs (N$_{tot.}$) with \feh\ 
estimates from each source and the final number (N$_{sel.}$) adopted in our 
spectroscopic catalog by following the quoted priority ranking.
The sky distribution of the final spectroscopic sample is plotted in
Fig.~\ref{fig:sky} where metallicity estimates coming from different sources 
are plotted with different symbols and/or colours.

In order to obtain a homogeneous spectroscopic catalog, we calibrated the
different literature \feh\ estimate based on LR datasets onto the same
metallicity scale used for the HR and \deltaS\ samples. Indeed, the "HR our"
(containing our own HR estimates), "HR lit" (containing the literature HR
estimates) and \deltaS\ estimates are already in the same metallicity scale.
Therefore we joined them into a single group (HR+\deltaS) of 7,997 RRLs and we
selected the RRLs in common between (HR+\deltaS) and the individual LR samples
(N$_{cal.}$). To convert the LR metallicities to our scale, we have fitted
$\feh_{HR+\Delta\rm S}$ as function of the $\feh_{LR}$, for each sample with
priority from 4 to 10 (see column 2 in Table~1). The coefficients of the
$\feh_{HR+\Delta\rm S} = a + b \cdot \feh_{LR}$ fits and their total
uncertainties are listed in Table~\ref{tbl:rrlcount}. Finally, we have adopted
the quoted fits to convert the metallicities from the LR samples into our scale.
Note that we could not perform this step for the RAVE metallicities because we
found only one match between the RAVE and the HR+\deltaS\ RRLs, therefore the
RAVE metallicities were not converted. We also note that, the fits for the
\citet{zinn20}, \citet{dambis13}, \citet{duffau14} and \citet{sesar13} samples
are close to the bisector, meaning that their metallicities are --- taking
account of the uncertainties --- already in a scale very similar to our own.
Additionally, we found that the metallicity scales of \citet{liu20} and
SEGUE-SSPP are different to ours. The difference between our own scale and that
of \citet{liu20} was already found by \citet{crestani21} and is due to the
different scale of their calibrators.

Note that a similar version of the current spectroscopic catalog, but only 
focussed on the radial velocity curve templates based on different spectroscopic 
diagnostics (metallic lines, Balmer lines), is discussed in the fifth paper of 
this series (Braga et al. 2021, ApJ submitted).

\begin{figure} 
\centering
\includegraphics[width=1\columnwidth]{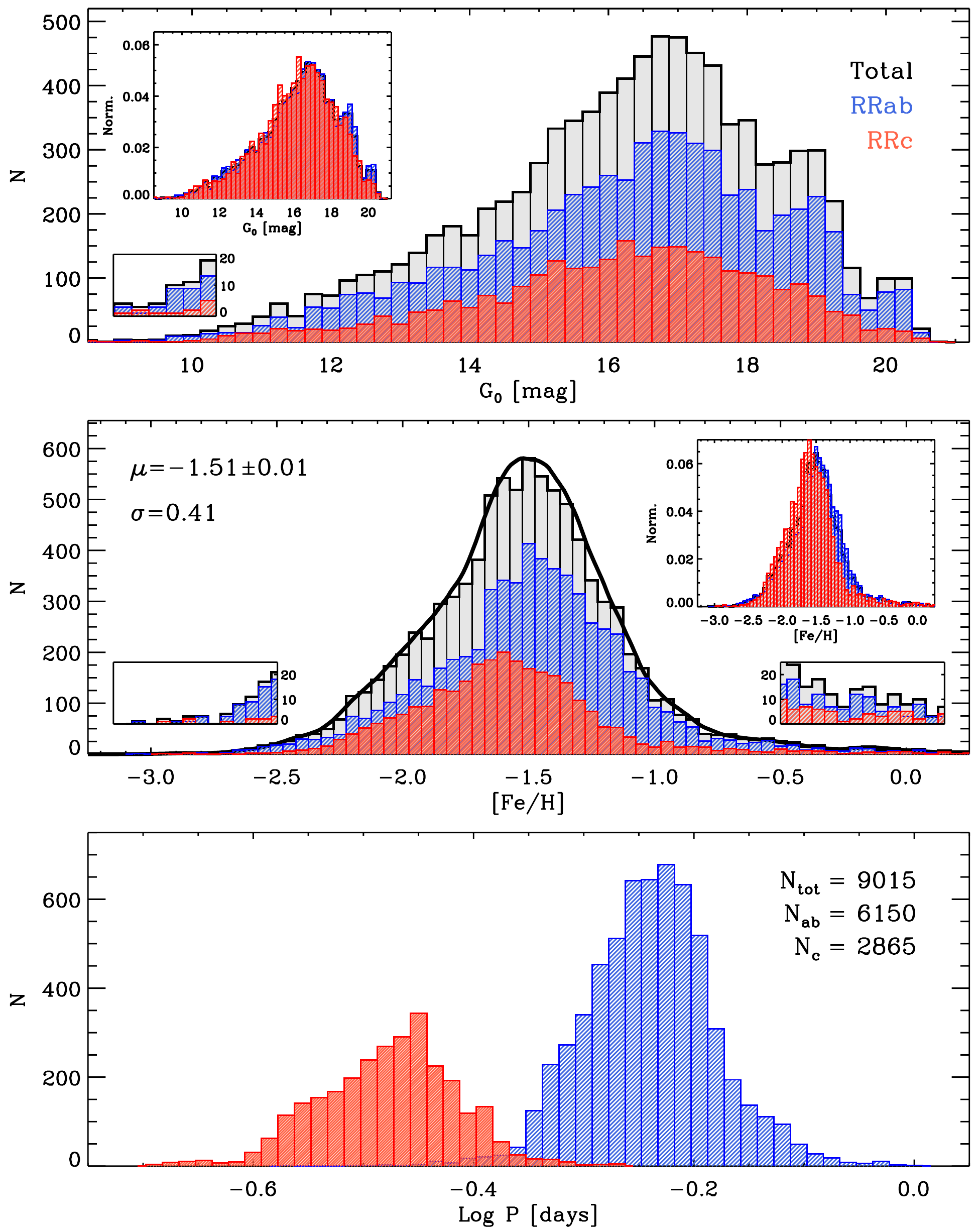} 
\caption{\textit{Top:} un-reddened \G\ magnitude distributions of the total RRL
sample (grey), RRab (blue) and RRc (red) samples with spectroscopic
measurements. The small inset displays a zoom-in of the histogram in the bright 
magnitude end. The large inset shows the same magnitude distributions, but area
normalised. 
\textit{Middle:} Metallicity distribution of the entire spectroscopic sample
(grey) together with the metallicity distribution for RRab (blue) and RRc (red)
variables. The solid line shows the smoothed metallicity distribution. The small
insets display zoom-in of the histogram in the metal-poor and metal-rich tails.
The large inset shows the same metallicity distributions, but area normalised.
\textit{Bottom:} Period distribution of the entire spectroscopic sample for RRab
(blue) and RRc (red) variables.}
\label{fig:distr}
\end{figure} 

\section{The metallicity distribution} \label{sec:iron}

Data plotted in the top panel of Fig.~\ref{fig:distr} display the apparent
un-reddened magnitude of the entire spectroscopic sample. The individual
reddening values were extracted from the \citet{schlegel98} dust maps and the
updated reddening coefficients from \citet{schlafly11}, while the extinction in
the $G$ band was calculated with the \citet{casagrande18} relation. The key
advantage of the current sample when compared with similar datasets available in
the literature is that RRLs cover the entire Halo, since their Galactocentric
distances range from a few~kpc to the outskirts of the Galactic Halo
(R$_G$$\ge$140~kpc). The individual distances were estimated by using predicted
optical, near-infrared and mid-infrared Period-Luminosity-Metallicity relations
\citep{marconi15,marconi18} and they will be discussed in a forthcoming paper.
The un-reddened apparent magnitude distributions of both RRab (blue) and RRc
(red) variables are quite similar, thus suggesting similar completeness limits.

The metallicity distribution of RRab variables is systematically more metal-rich
($\langle\feh\rangle_{ab}=-1.48\pm 0.01$, $\sigma=0.41$~dex) than that of the
RRc variables ($\langle\feh\rangle_{c}=-1.58\pm 0.01$, $\sigma=0.40$~dex, see
middle panel of Fig.~\ref{fig:distr}). This finding supports previous estimates
by \citet{liu20} and by \citet{crestani21}, which were based on smaller
datasets. In this context, it is worth mentioning that the RRc display a smooth
low-metallicity tail, while the RRab display a well defined jump for
$\feh\simeq-1.70$ followed by another small increase at $\feh\simeq-1.60$. In
fact, a metallicity peak for RRc more metal-poor than RRab variables is expected
from evidence on the distribution of horizontal branch (HB) stars across the RRL
instability strip. The current empirical and theoretical evidence indicates that
metallicity is the main parameter driving the HB morphology. An increase in the
metal content causes the HB morphology to become systematically redder
\citep{torelli19}. Stellar evolution theory \citep{bono19} and observations
\citep{coppola15,braga18} show that RRc populate the hottest and bluest portion
of the instability strip. The topology of the instability strip and the
dependence of the HB morphology on the metal content provide a qualitative
explanation of the reason why RRc variables can be more easily produced in the
metal-poor than in the metal-rich regime.

The bottom panel of Fig.~\ref{fig:distr} shows the period distribution of the
spectroscopic sample. Note that the current sample is covering the full period
range of RRLs. RRab have periods ranging from $\sim$0.4 days to almost one day,
while the RRc range from $\sim$0.2 days to $\sim$0.5 days, and the global
fraction of RRc variables is roughly 1/3 of the entire sample. This finding
supports theoretical predictions suggesting that the temperature region in which
RRc variables attain a stable pulsation cycle is roughly 1/3 of the entire width
in temperature of the instability strip \citep{bono94is}. Note that in this
plain explanation we are assuming that the central He burning lifetime of HB
stars is almost constant across the instability strip.
In passing, we note that the inclusion of RRc variables is crucial to
investigate the topology of the instability strip and to address several open
problems concerning field and cluster RRLs. However, their inclusion brings
forward the thorny problem of short period eclipsing binaries mimicking the
luminosity variation typical of RRc variables \citep{botan21}. To overcome the
contamination of eclipsing binaries we devised a new method based on the optical
\citep[\textit{Gaia},][]{clementini19} and mid-infrared \citep[MIR from
NEOWISE,][]{mainzer11} amplitude ratios. The eclipsing binaries, in the
amplitude ratio \textit{vs} pulsation period plane, cluster within the 
uncertainties around an amplitude ratio of 1, meaning that the MIR amplitude is 
similar to the the optical one. Regular RRL variables show in the same plane amplitude
ratios ranging from $\approx$0.2 (RRc) to $\approx$0.4 (RRab). However, the MIR
amplitudes are only available for $\sim$1\% of the RRc variables in the 
spectroscopic sample. To overcome this limitation we also used the Fourier 
parameters of the optical light curves together with a visual inspection of 
their light curves. This novel approach will be discussed in detail in a 
forthcoming paper (Mullen et al. 2021, in preparation).

\begin{figure*} 
\centering
\includegraphics[width=0.95\columnwidth]{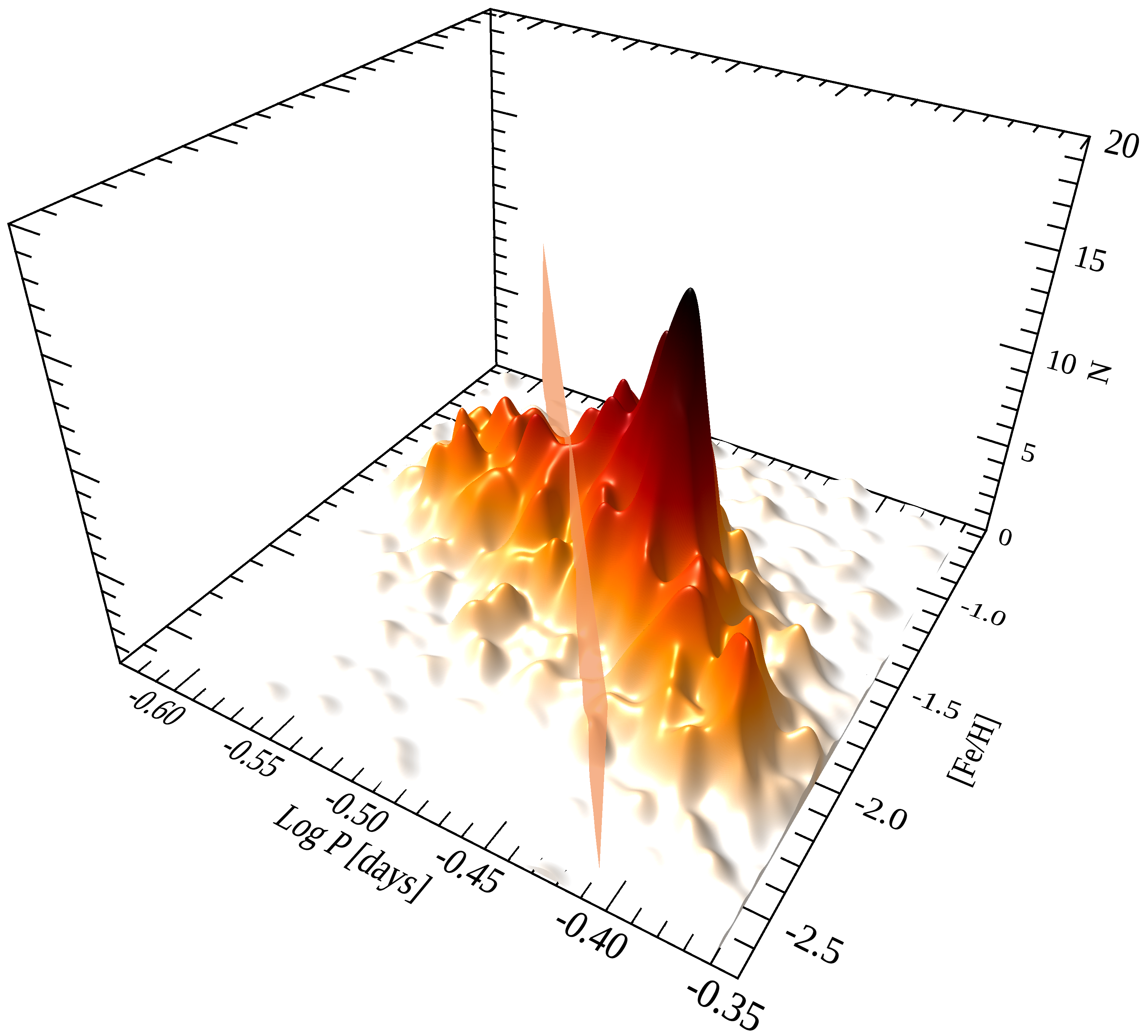} 
\hfill
\includegraphics[width=1\columnwidth]{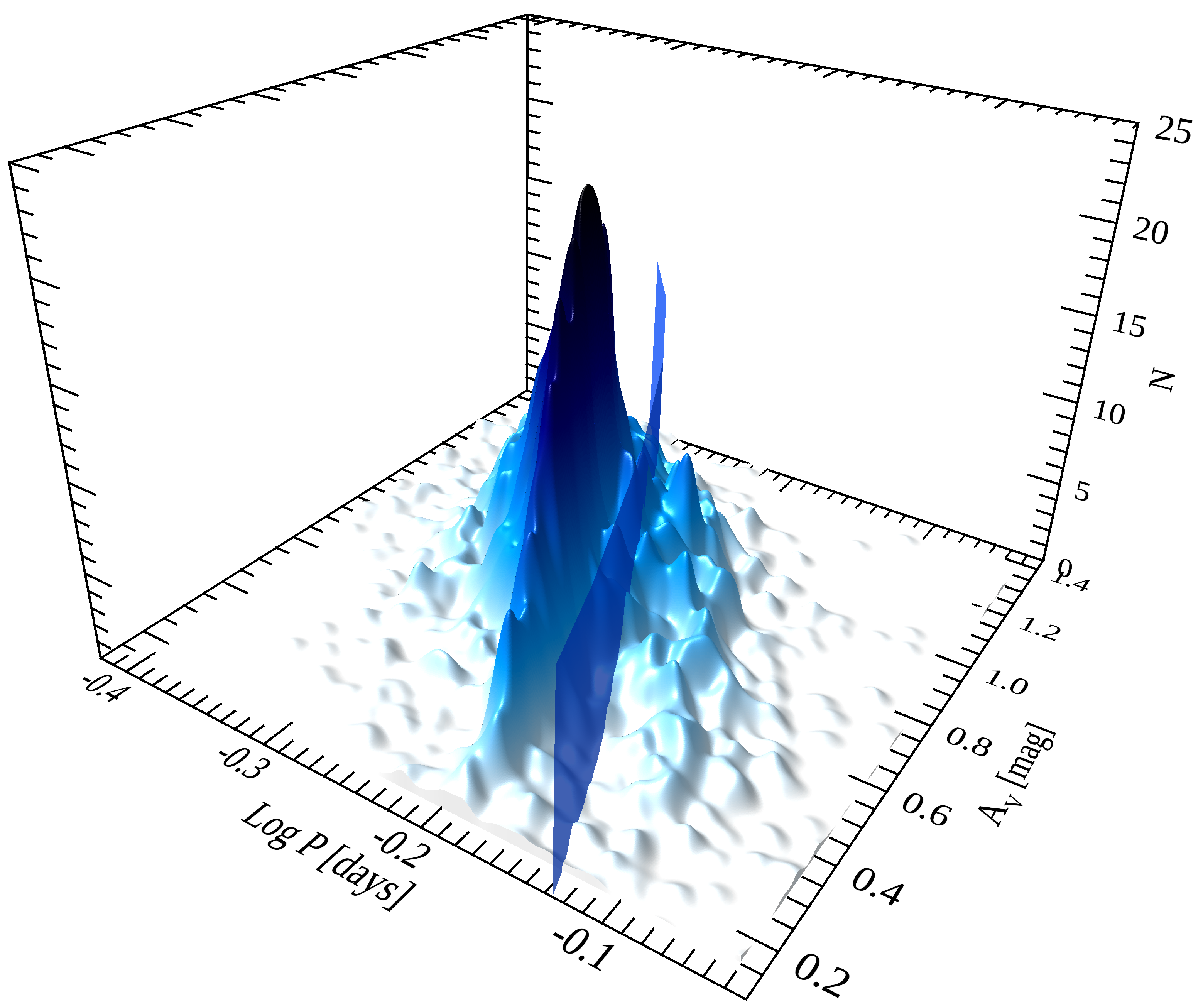} 
\caption{\textit{Left:} 3D period-metallicity distribution for RRc variables. The
distribution was smoothed by using a Gaussian kernel with unitary weight and $\sigma$
equal to the mean error on the metallicity. The light red plane splits the
short- from the long-period variables. 
\textit{Right:} same as the left, but for RRab variables in the 3D Bailey 
(logarithmic period \textit{vs} visual amplitude) distribution. The blue sky plane 
splits short from long-period RRab variables.}\label{fig:rrFits1}
\end{figure*} 

\begin{figure}[b] 
\centering
\includegraphics[width=1.\columnwidth]{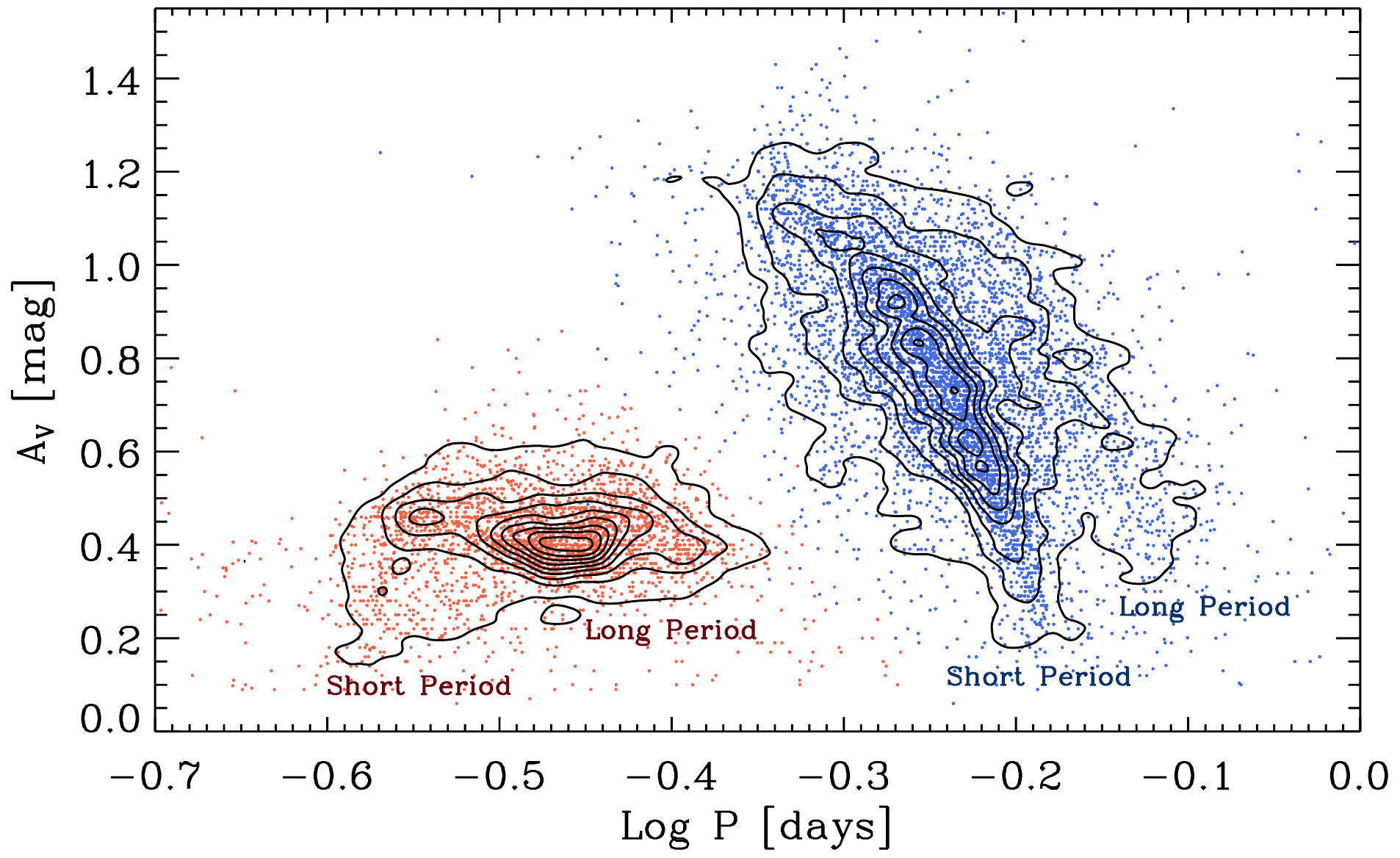} 
\caption{Bailey diagram of the spectroscopic sample: visual amplitude
\textit{vs} logarithmic period for both RRc (red) and RRab (blue) variables. The
contours display iso-density levels ranging from 5 to 95\% with steps of
$\sim$10\%.}\label{fig:rrCont}
\end{figure} 

\begin{figure*} 
\centering
\includegraphics[width=0.85\columnwidth]{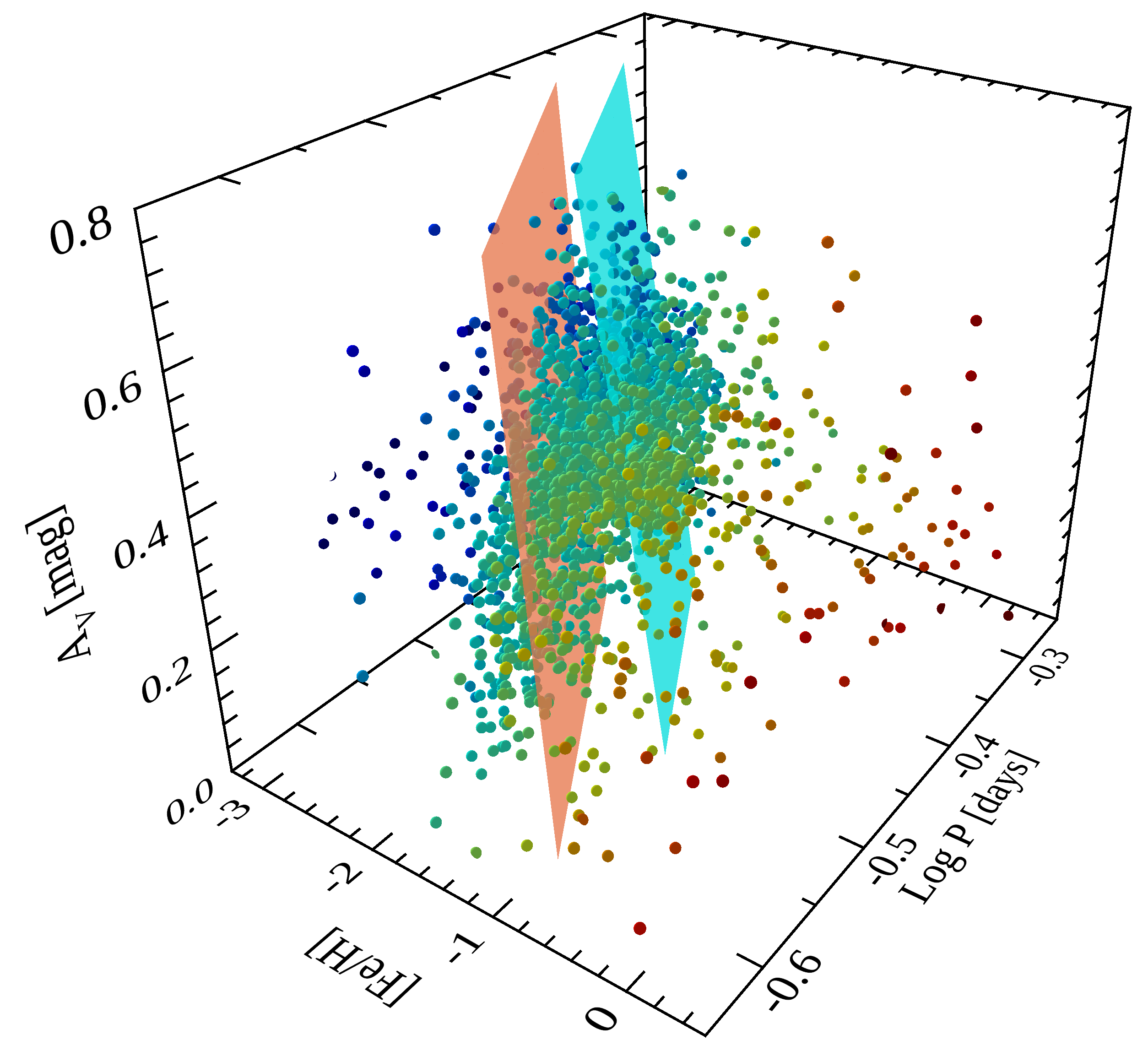} 
\includegraphics[width=1\columnwidth]{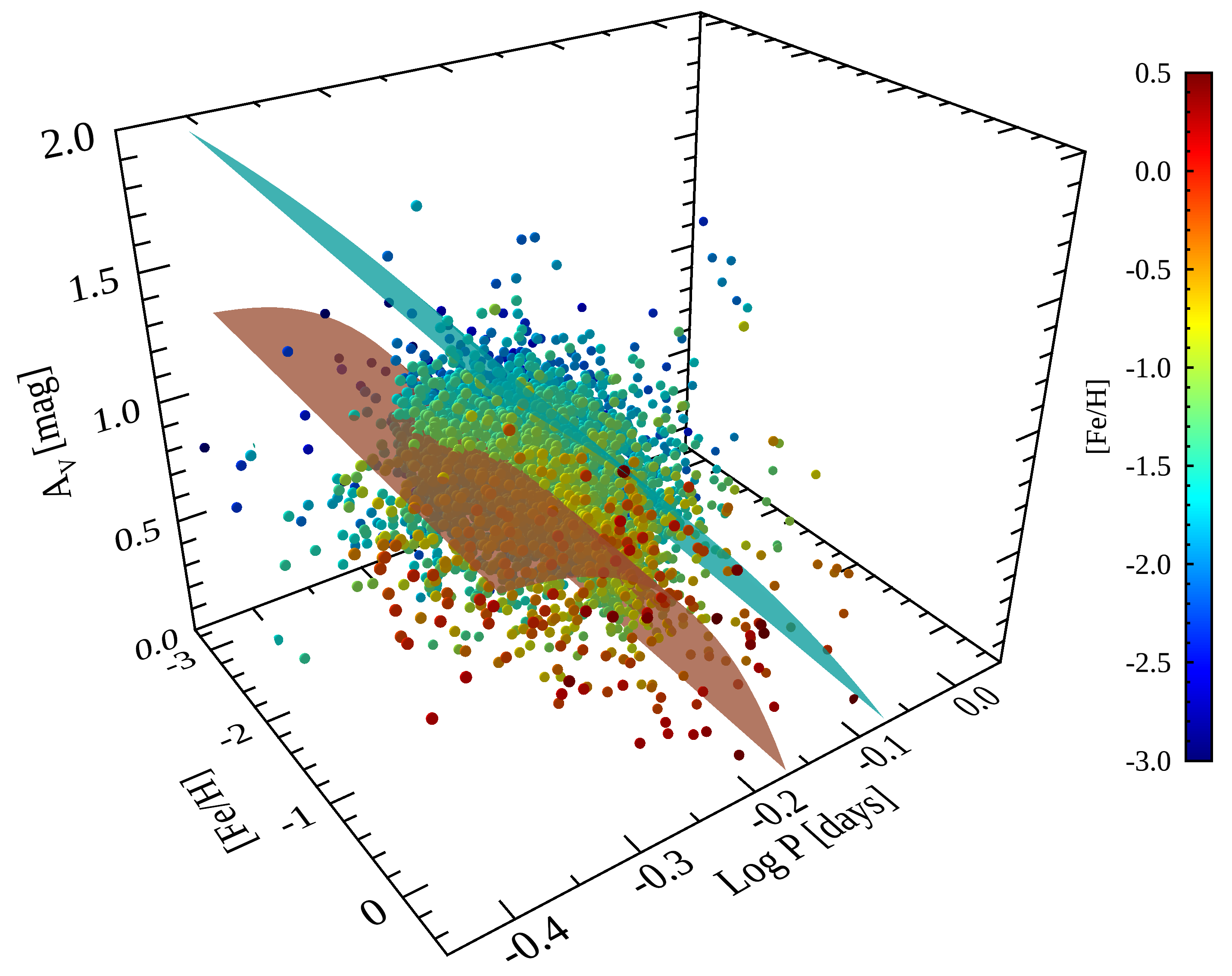} 
\caption{\textit{Left:} 3D Bailey diagram (visual amplitude, logarithmic period,
metallicity) for RRc variables. The metallicity is colour-coded  as shown by the 
colour bar on the right. The coral and cyan planes display the
analytical fits tracing more metal-rich (short-period) and more metal-poor 
(long-period) sub-groups of RRL variables (see text for more details). 
\textit{Right:} Same as the left, but for RRab variables.}\label{fig:rrFits}
\end{figure*} 

\section{The Bailey diagram} \label{sec:bailey}

\begin{figure*} 
\centering
\includegraphics[width=1.6\columnwidth]{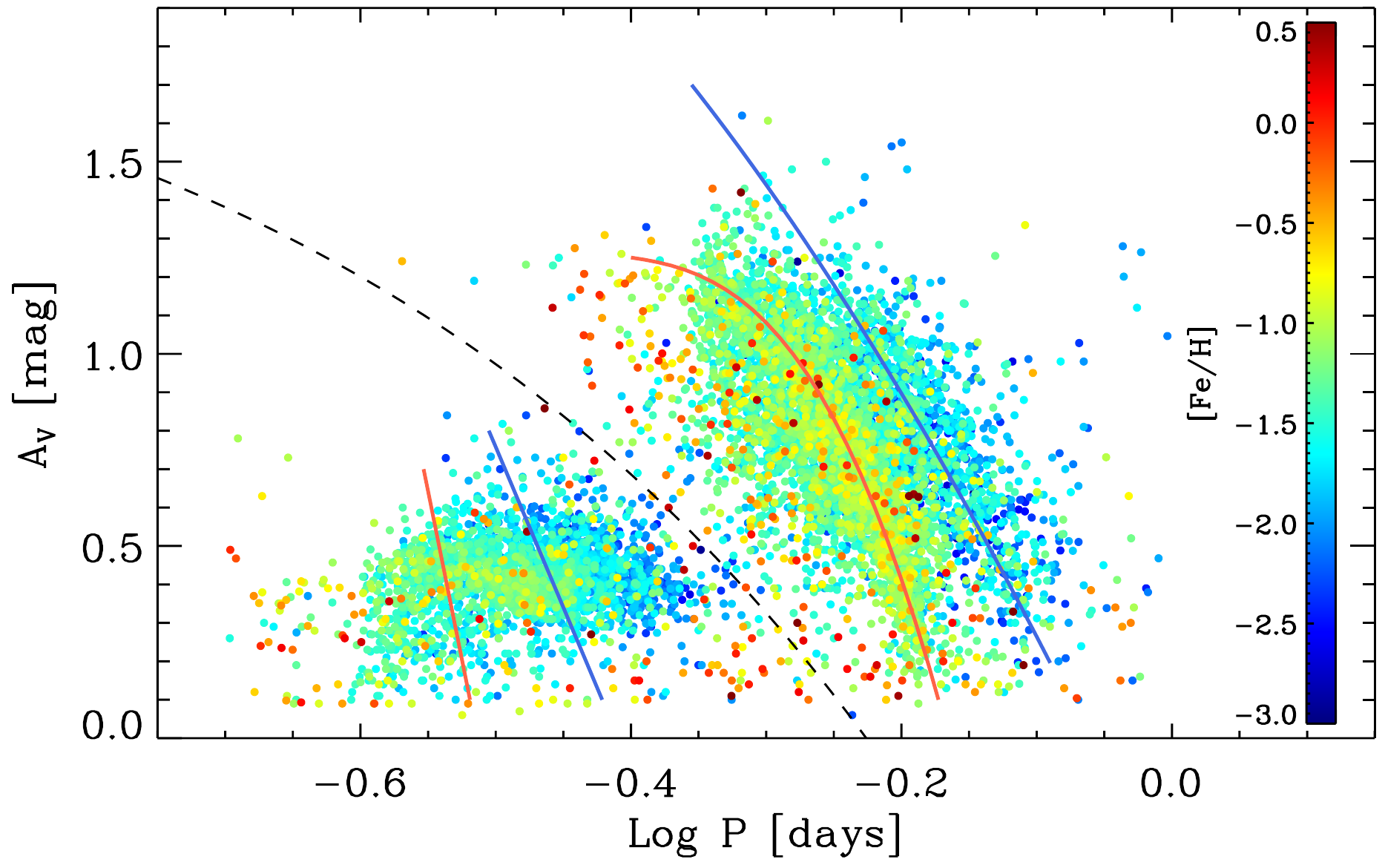} 
\caption{Bailey diagram of the spectroscopic sample. The metallicity is 
colour-coded as in Fig.~\ref{fig:rrFits} and the colour bar is plotted on the right.
The dashed line marks the relation used to separate RRc (shorter periods) 
and RRab (longer periods) variables \citet{clementini19}. The solid lines 
display the analytical relations tracing more metal-rich 
(short-period, red solid lines) and  
more metal-poor (long-period, blue solid lines) overdensities for both RRab and RRc variables 
(see text for more details).}
\label{fig:bailey}
\end{figure*} 

\begin{table*}
\caption{Parameters of the plane fits 
$A_V=a + b\cdot \feh + c\cdot\log P + d\cdot(\log P)^2 + e\cdot(\log P)^4$}
\label{tbl:ooCoeff}
\begin{tabular}{r c c c c c c}
\hline
\hline
 & $a\pm\epsilon_a$ & $b\pm\epsilon_b$ & $c\pm\epsilon_c$ & $d\pm\epsilon_d$ & $e\pm\epsilon_e$ & $\sigma_{\rm fit}$\\
\hline
RRab - Short Period & $-3.53\pm0.13$ & $-0.019\pm0.003$ & $-29.62\pm1.32$ &  $-52.37\pm3.63$ & $+49.88\pm7.69$ & 0.19\\
\hphantom{RRab - }Long Period &  $-0.51\pm0.06$ & $-0.032\pm0.007$ & $-7.69\pm0.68$ & $-4.52\pm1.83$ & \dots & 0.18 \\
RRc - Short Period & $-12.86\pm3.50$ & $-2.43\pm0.60$ & $-17.48\pm4.69$ & \dots & \dots & 0.77 \\
\hphantom{RRc -}Long Period & $-5.43\pm0.46$ & $-1.26\pm0.09$ & $-8.35\pm0.25$	 & \dots & \dots & 0.58\\
\hline
\end{tabular}
\end{table*}
      
To further improve the analysis of the fine structure of the Bailey diagram,
the left panel of Fig.~\ref{fig:rrFits1} shows in a 3D plot the distribution of
RRc variables in the logarithmic period \textit{vs} metallicity plane. The 3D
distribution was smoothed by using a Gaussian kernel with unitary weight and
$\sigma$ equal to the mean error on the metallicity estimates. The light red
plane%
\footnote{To separate SP and LP RRc variables we adopted the following plane: 
$\feh=-7.03(\pm 2.42)-10.80(\pm 3.88)\cdot\log P$ $[\sigma=0.82{\rm\,dex}]$} 
separate short-period (SP) from long-period (LP) RRc variables. The same
approach was adopted to separate short- from long-period RRab variables and the
right panel of Fig.~\ref{fig:rrFits1} shows the 3D plot of the Bailey diagram
(logarithmic period \textit{vs} visual amplitude). The 3D distribution was 
smoothed by using the same approach adopted for RRc variables, but the $\sigma$
is equal to the mean error on the luminosity amplitudes. The light blue 
plane%
\footnote{To separate SP and LP RRab variables we adopted the following plane: 
$A_V=-1.39(\pm 0.09)-13.76(\pm 0.97)\cdot\log P-15.10(\pm 2.64)\cdot\log P^2$ 
$[\sigma=0.19{\rm\,mag}]$}
separate SP from LP RRab variables. This plane is very similar to the
dotted–dashed line plotted in the bottom panel of Fig.~13 in \citet{fabrizio19}
showing the Oosterhoff intermediate loci and defined as the "valley" between the
two main overdensities. Note that the visual amplitudes ($A_V$) adopted in this
investigation come from two different sources:
\textit{a)} \G-band time series photometry collected by \textit{Gaia}. The light curves
were folded by using the periods provided within \textit{Gaia}~DR2 and fitted
with Fourier series. The \G-band amplitude was estimated as the difference
\G(min)--\G(max) of the analytical fit. The \G-band amplitudes were then
transformed into \vv-band amplitudes by using Eqn.2 from \citet{clementini19}.
\textit{b)} For RRLs with poor \textit{Gaia} phase coverage, $A_V$ estimates
were collected from the literature. As before, $A_V$ is the difference between
the brightest and faintest point of the analytical fit. It is worth mentioning
that, in the analysis of the Bailey diagram, we did not include RRLs for which
the mean magnitude was estimated by an optical light curve template \citep[i.e.
those from Pan-STARRS and DECam,][]{sesar17,stringer19}. In fact, despite
\G-band amplitudes are available for these stars, the amplitudes come from the
template fitting of the data, hence they are not homogeneous with $A_V$
estimates from the previous sources \textit{a)} and \textit{b)}.

The separation between SP and LP variables can be further investigated with
the iso-contour plots for both RRc (red dots) and RRab (blue dots) variables in
the canonical Bailey diagram of Fig.~\ref{fig:rrCont}. Data plotted in this
figure show that the distribution of RRLs is, as expected, far from being
homogeneous. The iso-contours associated with RRab variables show that the bulk
of RRab variables are mainly distributed along the SP sequence, while the LP
sequence only includes a minor fraction of RRab variables ($\sim$80\%
\textit{vs} $\sim$20\%). The RRc variables display an almost flat amplitude
distribution for periods ranging from $\log P=-0.57$ to $\log P=-0.35$. However,
the RRc variables display an opposite trend when compared with RRab variables,
indeed the fraction of SP ($\log P\leq -0.51$) RRc variables is significantly
smaller than the fraction of LP ones. The current data, indeed, are suggesting
relative fractions of $\sim$30\% and $\sim$70\%, respectively (see
Sect.~\ref{sec:ratio} for more quantitative details).

To summarise the observed correlation among pulsation period, visual
amplitude and iron content, Fig.~\ref{fig:rrFits} shows the 3D distribution of
the entire spectroscopic sample. Note that the iron content is colour-coded (see
the bar on the right) and moves from dark blue (very metal-poor) to dark red
(very metal-rich). 
We performed an analytical fit connecting the three key parameters (period,
visual amplitude, metallicity) independent of distance and reddening. The
coefficients of the fit, their errors and the standard deviations of the
different relations are listed in Table~\ref{tbl:ooCoeff}. Note that the cyan
and coral planes trace overdensities associated with more metal-rich (SP) and
more metal-poor (LP) sub-groups of RRL variables. In passing, we also note that
the standard deviation in metal content is, at a fixed pulsation period, too
large to apply these relations to individual RRLs. Indeed, the analytical
relations shall be applied to periods and amplitudes of sizeable RRL samples.

Fig.~\ref{fig:bailey} shows the classical Bailey diagram for the entire
spectroscopic sample, but the symbols are colour-coded following the same
metallicity scale adopted in Fig.~\ref{fig:rrFits}. To properly identify in the
canonical Bailey diagram the short- and the long-period sequences among the RRab
variables, the analytical fits discussed above were cut at fixed iron content
($\feh=-1.5$). The ensuing two-dimensional relations for the short- and the
long-period sequences are the following:    
\begin{eqnarray}
{\rm SP:\:} A_V[{\rm mag}] &=&  -3.50(\pm0.13)+\nonumber\\
\left[\sigma=0.19\:{\rm mag}\right]&& -29.62(\pm1.32)\cdot \log P+ \nonumber\\
&& -52.37(\pm3.63)\cdot(\log P)^2 +\nonumber\\
&& +49.88(\pm7.69)\cdot(\log P)^4\\
{\rm LP:\:} A_V[{\rm mag}] &=& -0.46(\pm0.06)+\nonumber \\
\left[\sigma=0.18\:{\rm mag}\right]&& -7.69(\pm0.68)\cdot \log P+ \nonumber \\
&& -4.52(\pm1.83)\cdot(\log P)^2
\end{eqnarray}
and they are plotted as a red and a blue solid line in the RRab region of
Fig.~\ref{fig:bailey}. 

We have already mentioned that RRc variables display a smooth transition when
moving from the short- to the long-period range. A plausible separation can only
be attained by using the 3D distribution. However, we decided to follow the same
approach adopted for RRab variables and we cut the analytical fits at
$\feh=-1.60$. The two dimensional analytical relations we obtained are the
following:   
\begin{eqnarray}
{\rm SP:\:} A_V[{\rm mag}] &=& -8.97(\pm3.55)+\nonumber\\
\left[\sigma=0.90\:{\rm mag}\right]&& -17.48(\pm4.69)\cdot \log P\\
{\rm LP:\:} A_V[{\rm mag}] &=& -3.42(\pm0.32)+\nonumber\\
\left[\sigma=0.40\:{\rm mag}\right]&& -8.35(\pm0.25)\cdot \log P 
\end{eqnarray}
and they are plotted as a red and a blue solid line in the RRc region of
Fig.~\ref{fig:bailey}.

Data plotted in this figure display quite clearly that an increase in the metal
content causes a systematic shift of both RRab and RRc towards shorter pulsation
periods. The adopted colour coding shows that the more metal-rich RRLs (from
yellow to red) are mainly located in the short-period tail. We also note that
the metallicity dependence is stronger among RRc than RRab variables. Indeed,
RRc with periods shorter than 0.25 days are systematically more metal-rich than
the bulk of RRc variables. A similar effect is also present among the RRab,
defining the so-called High Amplitude Short Period (HASP) variables
($P<0.48$~days, $A_V>0.75$~mag, \citealt{fiorentino15}). Indeed, the relative
fraction of RRab variables located in the HASP region more metal-rich than
$\feh=-1.5$ is 75\%, while their mean metallicity is $\langle\feh\rangle_{\rm
HASP}=-1.25\pm0.02$ ($\sigma=0.42$~dex). The current findings soundly support
the estimates by \citet{fabrizio19} using iron abundances based on
low-resolution spectra for $\sim$2,900 RRab variables and, more recently, by
\citet{crestani21} using iron abundances based on high-resolution spectra for
143 RRLs (111 RRab, 32 RRc). In passing, we also note that a similar trend for
the pulsation period as function of the metallicity was already found in the
Galactic Bulge by using data from the MACHO survey for a thousand of RRab stars
\citep{kunder09} and more recently by using more than 8,100 RRab variables
\citep{prudil19}.

\section{The dependence of periods and luminosity amplitudes on metallicity}
\label{sec:metalbins} 

\begin{figure*} 
\centering
\includegraphics[width=1.8\columnwidth]{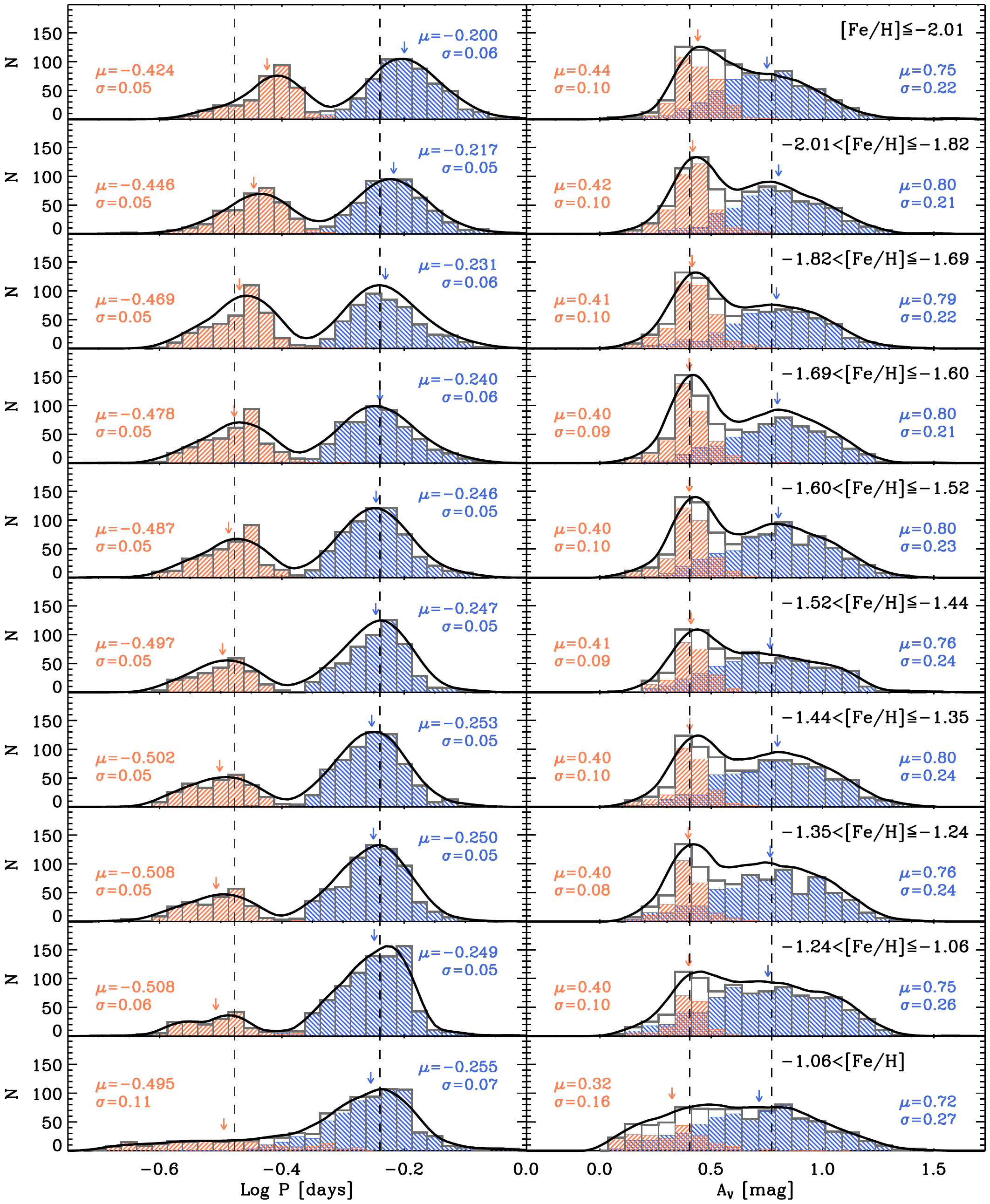} 
\caption{Period (\textit{left panels}) and visual amplitude (\textit{right
panels}) distributions of the spectroscopic sample. The sample was split in ten
metallicity bins (see labelled values) including a similar number of objects.
The solid lines display the smoothed distributions. The arrows mark the mean
period and the mean amplitude of the individual bins for RRab (blue) and RRc
(red) variables. The dashed lines show the mean $\log P$ ($-0.240$ and $-0.477$)
and mean $A_V$ (0.40 and 0.77~mag) of the total sample.}\label{fig:metal}
\end{figure*} 

To further investigate the dependence of both periods and amplitudes on the
metallicity we divided the entire spectroscopic sample into ten different
metallicity bins. The range in metallicity of the different bins was adjusted in
such a way that each bin includes one tenth of the total sample. The left 
and the right panels of Fig.~\ref{fig:metal} display the period and the visual 
amplitude distributions in the ten metallicity bins (see labeled values).

Data plotted in the left panels display very clearly that the mean of the period
distribution for both RRab (in blue) and RRc (in red) when moving from the metal-poor 
(top panels)
to the metal-rich (bottom panels) regime decreases from $\log P\simeq-0.20$ to
$\log P\simeq-0.26$ for RRab and from $\log P\simeq-0.42$ to $\log P\simeq-0.51$
for RRc variables. 
The dashed vertical lines show the mean periods of the entire sample
($\langle \log P\rangle_{ab}=-0.240$, $\langle \log P\rangle_{c}=-0.477$). The
variation of the mean period among the ten metallicity bins is of the order of
23\% ($\Delta \log P=0.055$) for RRab and of the order of 14\% ($\Delta \log
P=0.083$) for RRc variables.

The difference between the period distribution of RRab and RRc variables becomes
even more evident if we take into account the variation of the standard
deviations. Data plotted in the left panels show that the period distribution
becomes less and less peaked when moving from the metal-poor to the metal-rich
regime. This trend is more relevant for RRc variables, because the standard
deviation in the most metal-rich bin is a factor of two larger than the typical
standard deviation of the metal-poor metallicity bins, despite all bins sharing
the same number of objects, and although the last bin covers a large range in 
metallicity.  

The visual amplitudes display a similar behaviour (right panels): an increase in
metal content causes a steady decrease in the mean visual amplitude.
The effect is more relevant for RRc than for RRab variables. Indeed, the A$_V$
for RRc variables decreases from $\sim$0.44~mag to 0.32~mag ($\sim$25\%). The
RRab variables show a different trend: the amplitude increases from
$\sim$0.75~mag to $\sim$0.80~mag, when moving from the metal-poor to the
metal-intermediate regime, while it decreases from $\sim$0.80~mag to
$\sim$0.72~mag when moving from the metal-intermediate to the metal-rich regime.

\begin{figure} 
\centering
\includegraphics[width=1.\columnwidth]{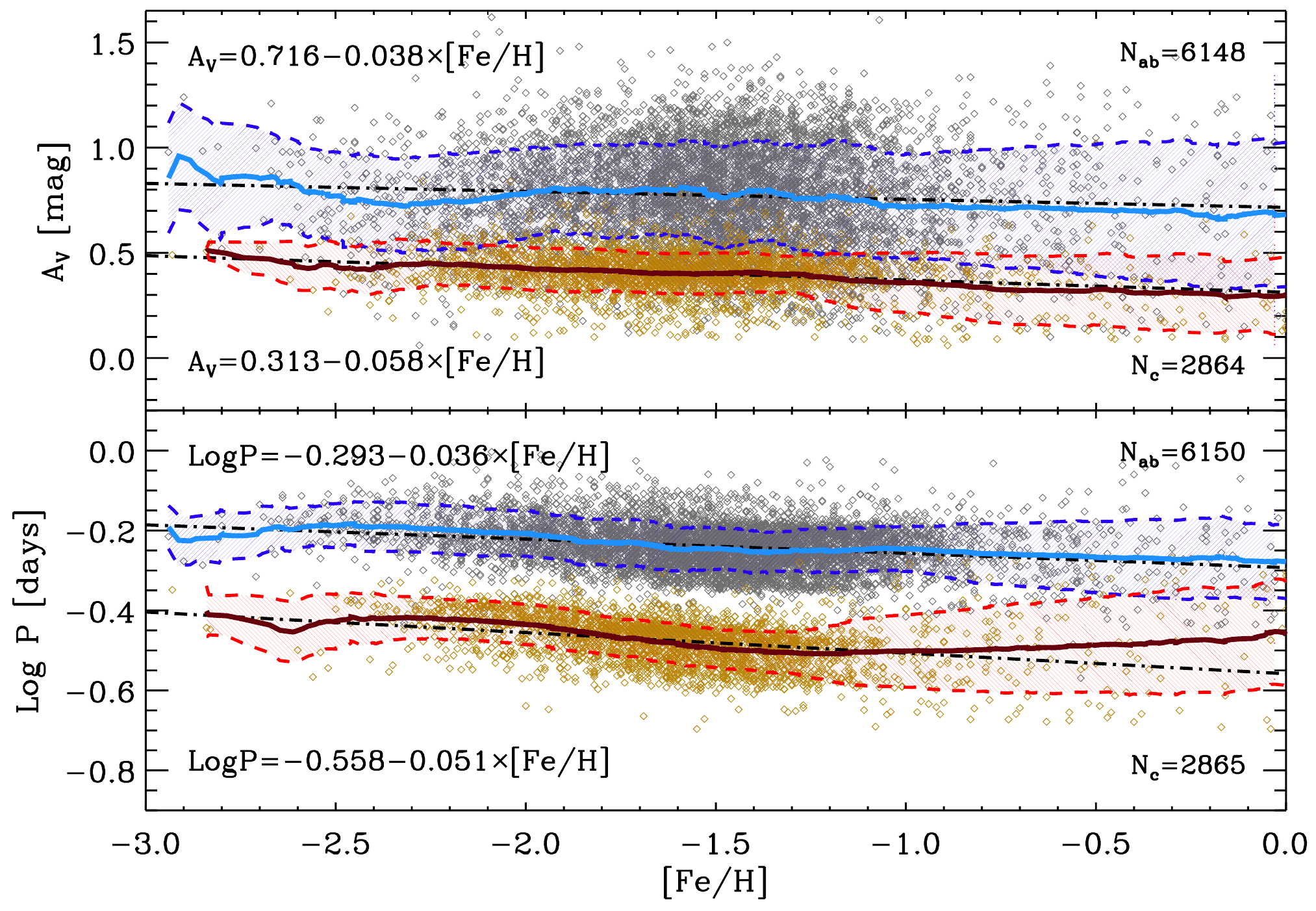} 
\caption{\textit{Top:} Visual amplitude as function of iron abundance for RRab
(grey dots) and RRc (dark yellow dots). The blue and the red solid lines display
the running averages for RRab and RRc variables and the shaded areas show 
the corresponding standard deviation. The dot-dashed lines
display the linear regressions and their coefficients are labelled. 
\textit{Bottom:} Same as the top, but with the $\log P$ on the y-axis.}
\label{fig:fits}
\end{figure} 

\begin{figure*} 
\centering
\includegraphics[width=1.5\columnwidth]{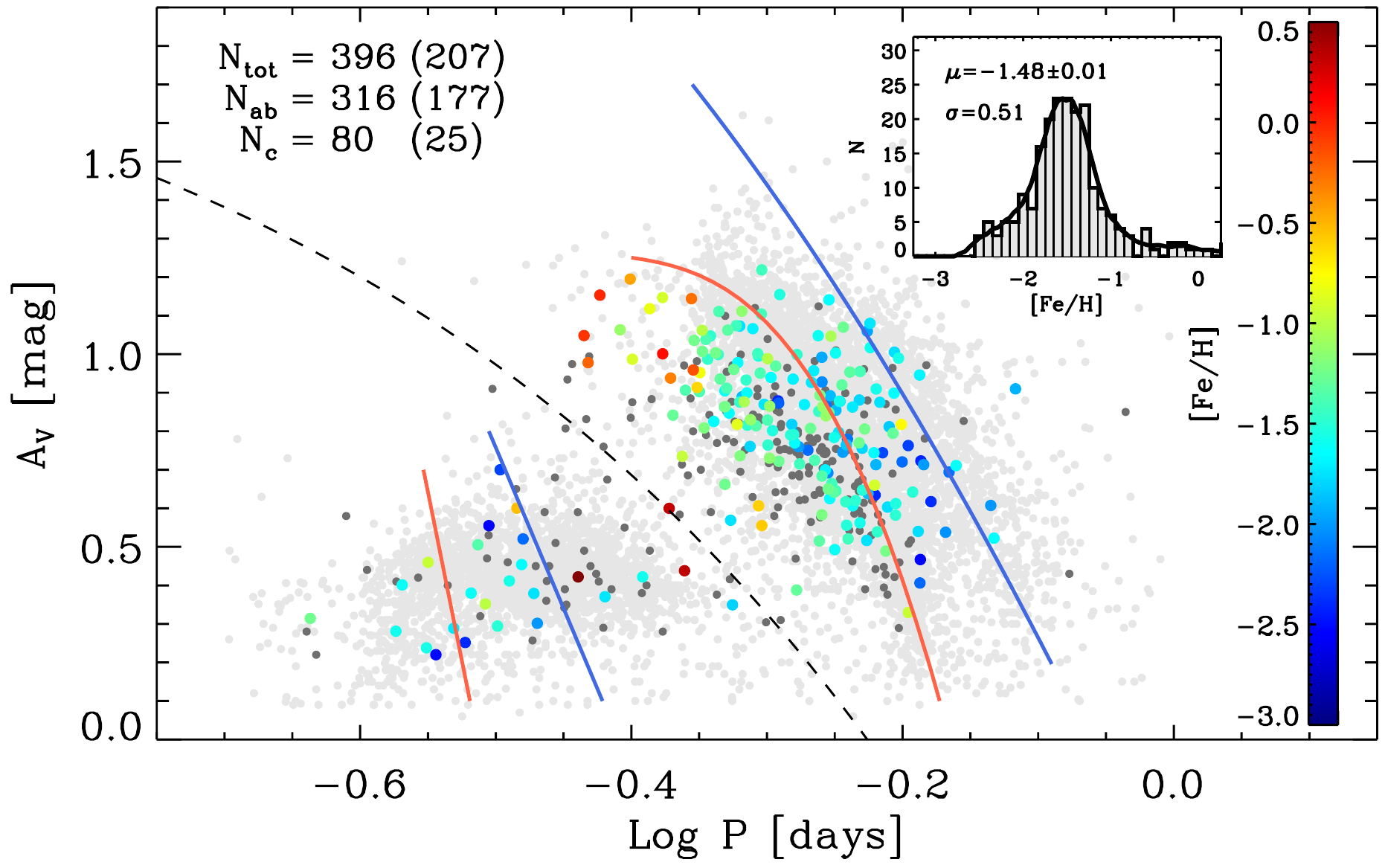} 
\caption{Bailey diagram of candidate Blazhko RRLs. The light grey symbols
display the entire spectroscopic sample plotted in Fig.~\ref{fig:bailey}. The
Blazhko RRLs for which the iron abundance is available are marked with coloured
symbols, while those for which it is not available are marked with dark grey
symbols. The numbers of RRab and RRc candidate Blazhko RRLs are labelled, the
numbers in parentheses display the numbers of candidate Blazhko RRLs with iron
abundances. The small inset shows the metallicity distribution of candidate 
Blazhko RRLs.}\label{fig:blazhko}
\end{figure*} 

To investigate on a more quantitative basis the metallicity dependence of both
periods and visual amplitudes we performed a linear fit over the entire
metallicity range (see Fig.~\ref{fig:fits}). To overcome spurious fluctuations
in the metal-poor and in the metal-rich regime we performed a running average.
The entire sample of RRab variables (grey diamonds) was ranked as a function of
the metal content, and we estimated the mean with a running box including 500
variables. The solid blue line plotted in the top panel of Fig.~\ref{fig:fits}
shows the running average and the two dashed lines display the 1$\sigma$
standard deviation. The same approach was adopted for RRc variables (yellow
diamonds), the red solid and dashed lines plotted in the same panel show their
running average and their standard deviations.

To validate the spectroscopic sample adopted to estimate the dependence of 
the luminosity amplitude on the metal content we performed several tests. To
quantify the impact that RRL with amplitude modulation have on the global
distribution we neglected candidate Blazhko RRLs (see Sect.~\ref{sec:blazhko}).
Furthermore, we also tested the dependence on the faint tail and we neglected 
variables fainter than the \G=17~mag (i.e. the peak in the apparent magnitude 
distribution of Fig.~\ref{fig:distr}). We found that the coefficients of the 
analytical relations for both the visual amplitudes and the pulsation periods 
are, within the errors, identical.

To constrain on a more quantitative basis the difference between the trend of RRab 
and RRc variables we performed a linear fit of the entire sample and we found: 
\begin{eqnarray}
{\rm RRab:\:} A_V[{\rm mag}]&=&0.716(\pm 0.011)+\nonumber\\
\left[\sigma=0.24\right]  \;\; && -0.038(\pm 0.007) \cdot \feh
\end{eqnarray}
and 
\begin{eqnarray}
{\rm RRc:\:} A_V[{\rm mag}]&=&0.312(\pm 0.008)+\nonumber\\
\left[\sigma=0.10\right] && -0.058(\pm 0.004) \cdot \feh
\end{eqnarray}

Note that the number of RRLs adopted in the analytical fits of the visual
amplitude as a function of the metal content is smaller than the number of RRLs
adopted in analytical fits of the pulsation periods because in the former sample 
we neglected the RRLs whose mean magnitude was estimated by using an optical 
light curve template, i.e. variables for which the visual amplitude is not 
available yet.
The former relation soundly supports the results obtained by \citet{fabrizio19}
for a smaller sample of RRLs (2,903 \textit{vs} 6,150 RRab). 
Moreover, the current amplitude-metallicity relation for RRab variables 
further supports the modest dependence of the visual amplitude on the metallicity. 
Indeed, a variation of 3~dex in metal content would only cause a difference 
in visual amplitude of $\approx$0.12~mag. The metallicity dependence is stronger 
for RRc variables: an increase of 3~dex in metal content causes an increase in 
visual amplitude that is almost a factor of two larger ($\approx$0.2~mag).
The stronger sensitivity of RRc visual amplitudes to metallicity and the decrease 
by a factor of two in the standard deviation (0.10 \textit{vs} 0.24~mag) are due to 
the fact that the region of the instability strip in which they are pulsationally 
stable is at least a factor of two narrower than the region covered by RRab 
variables.
This means that the impact of the metallicity on
pulsation and evolutionary properties for RRc variables is less affected by
changes in their intrinsic properties. Evolutionary effects due to off-Zero Age
HB (ZAHB) evolution at fixed stellar mass and chemical composition and for a
typical red-ward evolution \citep[see Fig. 4 in][]{bono20}, causes a decrease in
surface gravity, and in turn, an increase in the pulsation period and a decrease
in the visual amplitude. This means that the off-ZAHB evolution can have a
complex pattern across the Bailey diagram \citep{bono20}.

The bottom panel of Fig.~\ref{fig:fits} shows the logarithmic period of the
entire spectroscopic sample as a function of the metal content. We estimated the
same running averages (blue and red solid lines) over the two subsamples and
performed the linear fits (dot-dashed lines) obtaining the following results:
\begin{eqnarray}
{\rm RRab:\:} \log P [{\rm days}] &=& -0.293(\pm 0.003)+\nonumber\\
\left[\sigma=0.06\right] && -0.036(\pm 0.002) \cdot \feh
\end{eqnarray}
and 
\begin{eqnarray}
{\rm RRc:\:} \log P [{\rm days}] &=& -0.558(\pm 0.004)+\nonumber\\
\left[\sigma=0.06\right] && -0.051(\pm 0.003) \cdot \feh
\end{eqnarray}
    
As expected, the dependence of the pulsation period for both RRab and RRc variables 
on the metallicity are similar to the dependence on the visual amplitude, but their 
standard deviations are systematically smaller. Moreover, preliminary
evidence indicates that the dispersion of the $\log P$-metallicity relation for
RRc variables in the metal-rich regime is larger than for RRab. In this context
it is worth mentioning that the two sub-samples are well populated for
$\feh\ge-0.7$ (218 RRab; 90 RRc). We still lack a quantitative explanation for
this effect. The global properties of RRab and RRc variables will be addressed
in the next section.

\subsection{Physical mechanisms affecting the Bailey diagram} 
\label{sec:blazhko}

\begin{figure} 
\centering
\includegraphics[width=1.\columnwidth]{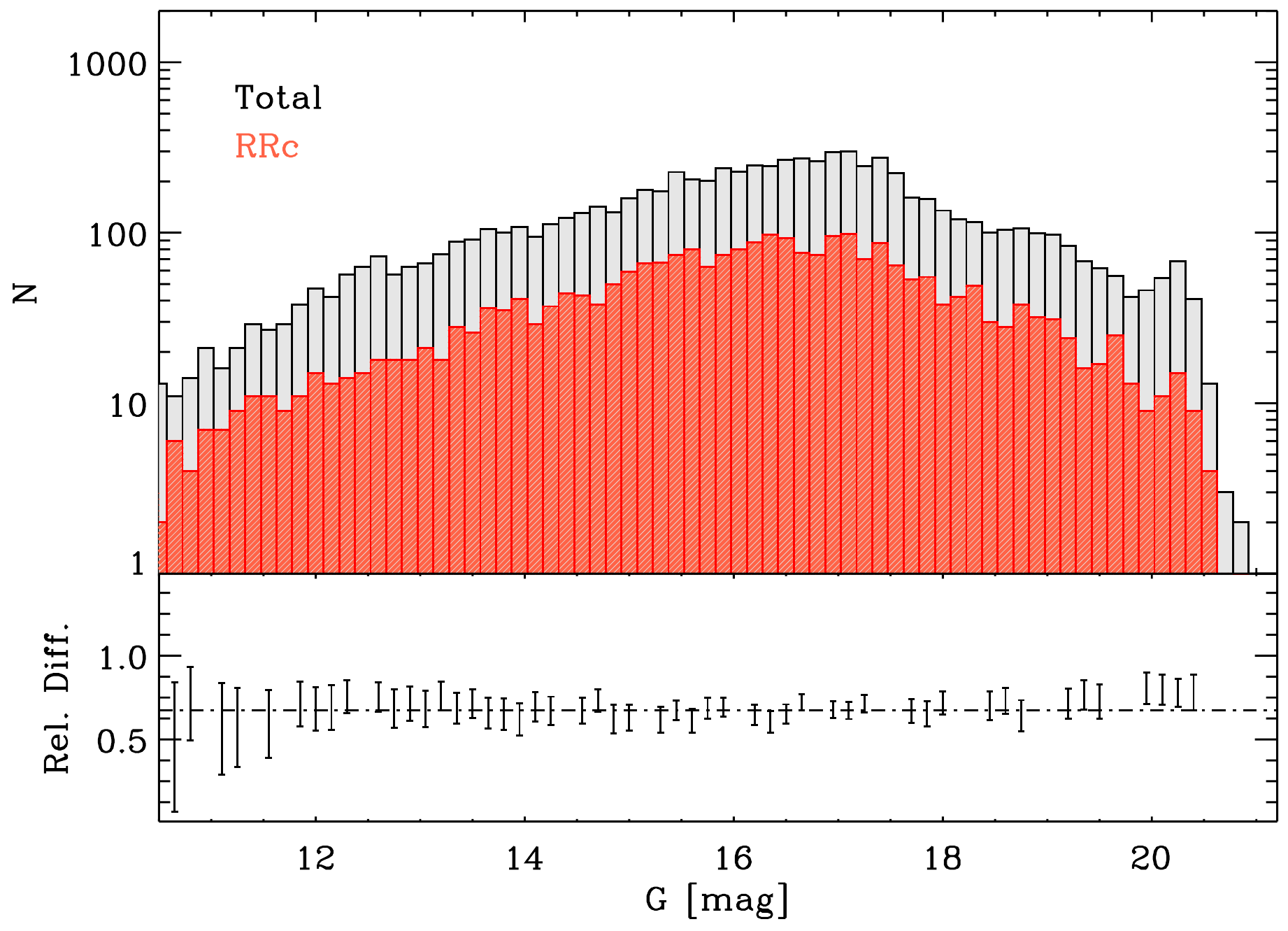} 
\caption{\textit{Top:} Apparent \G-band distribution of RRLs in the spectroscopic
sample. The red histogram shows the distribution of RRc variables, while the grey one
the global (RRab+RRc) sample. \textit{Bottom:} Relative difference between RRc 
distribution and the global sample.}\label{fig:Gratio}
\end{figure} 

\begin{figure*} 
\centering
\includegraphics[width=1.5\columnwidth]{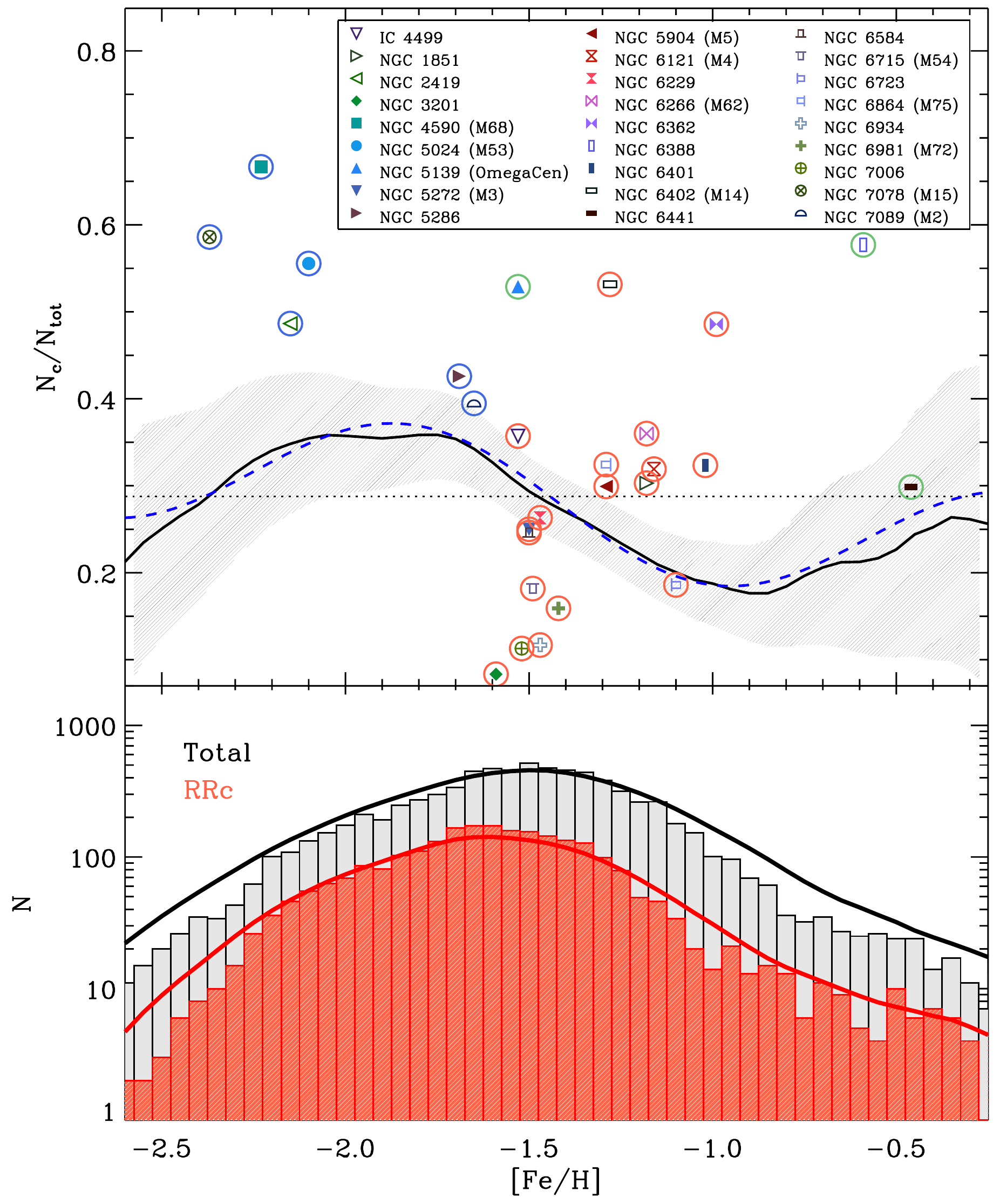} 
\caption{\textit{Top:} The population ratio as a function of the iron abundance
(solid black line), i.e. the ratio between the number of RRc and the total
number ($N_c/(N_c+N_{ab})$) of RRL variables. The grey hatched area shows the
1$\sigma$ uncertainty. Coloured symbols display the population ratio of Galactic
GCs (see labelled names) hosting at least 33 RRLs. The GCs hosting RRLs with
pulsation properties typical of OoI stellar systems are marked with an open red
circle, while those with pulsation properties typical of OoII systems are marked
with an open blue circle. Three peculiars clusters (NGC~6388, NGC~6441,
$\omega$~Cen) are marked with an open green circle. The blue dashed line shows
the analytical fit of Eqn.\ref{eqn:sinfit}.
\textit{Bottom:} Iron distribution for RRc (red) and total (RRab+RRc, grey) RRL
variables. The solid lines show the same distribution, but smoothed using a
Gaussian kernel with unitary weight and $\sigma$ equal to the individual errors
on the iron abundances.}\label{fig:ratio}
\end{figure*} 

Together with the evolutionary effects already mentioned in
Sect.~\ref{sec:metalbins} the distribution  of the variables in the Bailey
diagram is also affected by two independent physical mechanisms: \textit{a)} the
Blazhko effect and \textit{b)} non-linear phenomena. The Blazhko phenomenon
mainly affects the RRab variables \citep[$\sim$40\%,][]{prudil17} and the
modulation in amplitude is more relevant for shorter than for longer period
RRLs. Indeed, the modulations change from $\sim$0.7~mag, close to the
fundamental blue edge, to $\sim$0.05~mag, close to the fundamental red edge (see
Fig.~10 in \citealt{skarka20}, and also \citealt{jurcsik11,benko14,braga16}).
The RRc variables are also affected by the Blazhko phenomenon, but the fraction
is significantly smaller \citep[$\sim$6\%,][]{netzel18} and the amplitude
modulation is, at most, of the order of $\sim$0.2~mag.

To further investigate the impact that the Blazhko effect has on the trends
visible in the Bailey diagram, Fig.~\ref{fig:blazhko} shows the candidate
Blazhko Halo RRLs\footnote{\url{https://www.physics.muni.cz/~blasgalf/}}. They
are plotted using their mean visual amplitude and we are showing the entire
sample of candidate Blazhko RRLs, i.e. RRLs with amplitude modulations ranging
from a few hundredths to a few tenths of magnitude. The candidate Blazhko RRLs
for which the iron abundance is available are marked with coloured symbols (see
the bar on the right side), while those lacking of a metallicity estimate are
plotted with dark grey symbols. 
The metallicity distribution of candidate Blazhko RRLs (see the inset in
Fig.~\ref{fig:blazhko}) is quite similar to the global RRL metallicity
distribution. Although, the sample of candidate Blazhko RRLs with iron abundances
is at least forty times smaller of the entire spectroscopic sample, both the
peak and the standard deviation agree within the errors. We cannot exclude
possible biases in the metallicity distribution, because the candidate Blazhko
RRLs, as noted by the anonymous referee, are mainly restricted to the solar
neighbourhood. However, the tails of the RRab metallicity distribution appear to
be properly sampled, while for RRc variables the metallicity distribution is
still too limited. Data plotted in the Bailey diagram display that the Blazhko
phenomenon for RRab variables appears to be mainly associated either with
metal-intermediate or with metal-rich RRLs. Indeed, in agreement with the global
trend (see Sect.~\ref{sec:bailey}) the bulk of the candidate Blazhko RRab
variables are mainly distributed across the SP sequence (92\% \textit{vs} 8\%
across the LP sequence), while the candidate Blazhko RRc are mostly located
across the LP sequence (78\% \textit{vs} 22\% across the SP sequence). The
relative fractions for RRc variables need to be cautiously treated, since they
are roughly two dozen.
In passing, we also note that the current
findings support recent results by \citet{skarka20} based on a large sample
(more than 3,300) of Galactic Bulge and cluster Blazhko RRLs.

Non-linear phenomena like the formation and propagation of strong shocks are
mainly affecting RRab variables and, in particular, the RRab located close to
the blue edge of the instability strip. These variables are characterised by
very large amplitudes and light curves showing a saw-tooth shape. The occurrence
of non-linear phenomena is strongly supported by the presence of a dip along the
rising branch of the light curve and by solid spectroscopic evidence
\citep{gillet13,sneden17,gillet19}. The RRc variables are also affected by
shocks, but the current evidence indicate that the shocks marginally affect
their properties \citep{duan20,benko21}.

The current circumstantial evidence indicates that the Blazhko effect and
non-linear phenomena can only partially account for the typical dispersion of
the luminosity amplitudes, at fixed pulsation period, and in particular, the
difference between the RRab and the RRc when these parameters are considered.

\begin{figure*} 
\centering
\includegraphics[width=1.3\columnwidth]{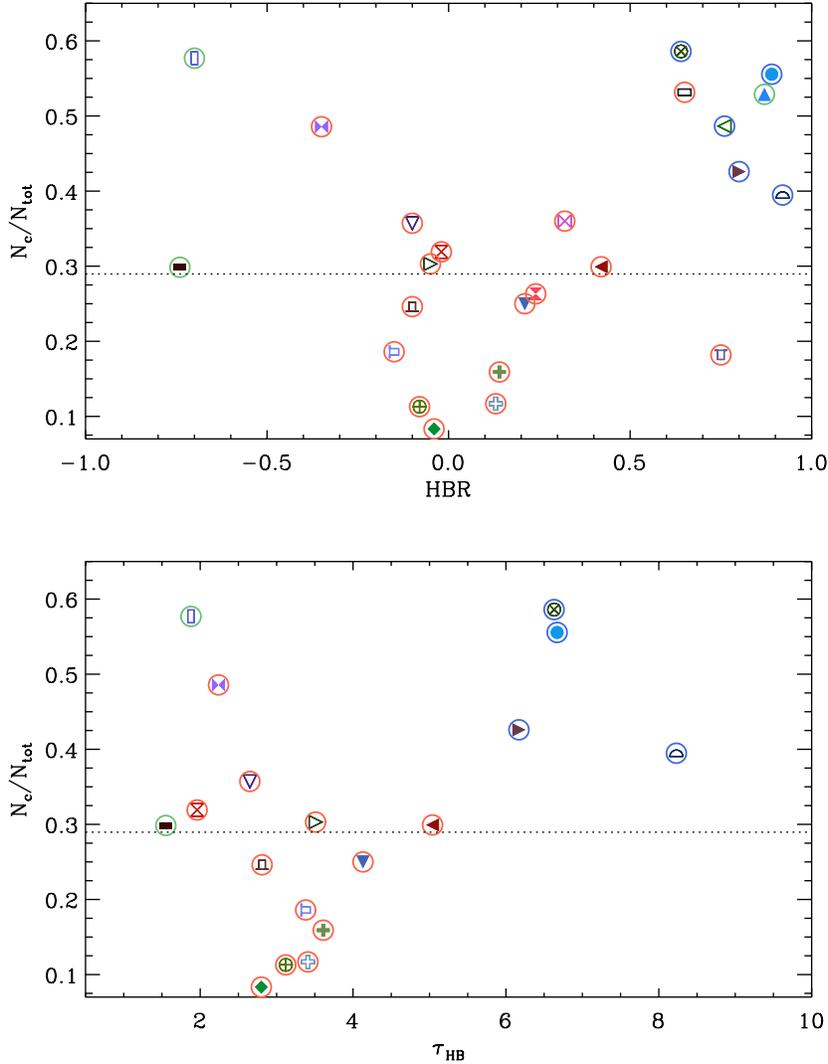} 
\caption{\textit{Top:} The population ratio N$_c$/N$_{tot}$ as function of the
HBR parameter. The symbols are the same as Fig.~\ref{fig:ratio}.
\textit{Bottom:} Same as the top, but as function of $\tau_{\rm HB}$
parameter.}\label{fig:ratio3}
\end{figure*} 

\section{The dependence of the population ratio on metallicity} 
\label{sec:ratio}

\begin{table*}
\caption{Galactic Globular Clusters hosting more than 33 RR Lyrae stars. From
left to right the different columns give the cluster name(s), the iron
abundance, the number of RRab and RRc variables, the population ratio, the
horizontal branch morphology indices and the Oosterhoff type (see text for more
details).}
\label{tbl:ggc}
\centering
\begin{tabular}{l r r r r r r r}
\hline
\hline
Cluster & \feh$^a$ & N$_{ab}$$^b$ & N$_{c}$$^b$ & N$_{c}$/N$_{tot}$ & $HBR$$^c$ & $\tau_{HB}$$^c$ & Ooster.Type\\
\hline
IC 4499& $-1.53$&          63&          35& 0.36&$-0.10$& 2.65&OoI\\
NGC 1851&$-1.18$&          23&          10& 0.30&$-0.05$& 3.51&OoI\\
NGC 2419&$-2.15$&          38&          36& 0.49& 0.76&\dots&OoII\\
NGC 3201&$-1.59$&          77&           7& 0.08&$-0.04$& 2.80&OoI\\
NGC 4590 (M68)&$-2.23$&          14&          28& 0.67& 0.58& 4.43&OoII\\
NGC 5024 (M53)&$-2.10$&          28&          35& 0.56& 0.89& 6.67&OoII\\
NGC 5139 ($\omega$ Cen)&$-1.53$$^e$&          90&        101& 0.53& 0.87&\dots&Pecul.$^d$\\
NGC 5272 (M3)&$-1.50$&         177&          59& 0.25& 0.21& 4.13&OoI\\
NGC 5286&$-1.69$&          31&          23& 0.43& 0.80& 6.17&OoII\\
NGC 5904 (M5)&$-1.29$&          89&          38& 0.30& 0.42& 5.04&OoI\\
NGC 6121 (M4)&$-1.16$&          32&          15& 0.32&$-0.02$& 1.96&OoI\\
NGC 6229&$-1.47$&          42&          15& 0.26& 0.24&\dots&OoI\\
NGC 6266 (M62)&$-1.18$&         144&          81& 0.36& 0.32&\dots&OoI\\
NGC 6362&$-0.99$&          18&          17& 0.49&$-0.35$& 2.24&OoI\\
NGC 6388&$-0.59$&          11&          15& 0.58&$-0.70$& 1.88&Pecul.$^d$\\
NGC 6401&$-1.02$&          23&          11& 0.32&\dots&\dots&OoI\\
NGC 6402 (M14)&$-1.28$&          52&          59& 0.53& 0.65&\dots&OoI\\
NGC 6441&$-0.46$&          47&          20& 0.30&$-0.74$& 1.55&Pecul.$^d$\\
NGC 6584&$-1.50$&          49&          16& 0.25&$-0.10$& 2.81&OoI\\
NGC 6715 (M54)&$-1.49$&         153&          34& 0.18& 0.75&\dots&OoI\\
NGC 6723&$-1.10$&          35&           8& 0.19&$-0.15$& 3.38&OoI\\
NGC 6864 (M75)&$-1.29$&          25&          12& 0.32&\dots&\dots&OoI\\
NGC 6934&$-1.47$&          68&           9& 0.12& 0.13& 3.41&OoI\\
NGC 6981 (M72)&$-1.42$&          37&           7& 0.16& 0.14& 3.61&OoI\\
NGC 7006&$-1.52$&          55&           7& 0.11&$-0.08$& 3.12&OoI\\
NGC 7078 (M15)&$-2.37$&          65&          92& 0.59& 0.64& 6.63&OoII\\
NGC 7089 (M2)&$-1.65$&          23&          15& 0.39& 0.92& 8.23&OoII\\
     \hline
\end{tabular}
\tablenotetext{}{$^a$\citet{harris96}. $^b$\citet{clement01}.
$^c$\citet{torelli19}. $^d$\citet{braga16}. $^e$The metallicity range covered by
RRLs is $-2.58\le\feh\le-0.85$ on the basis of spectroscopic measurements
\citep{magurno19} and photometric indices \citep{bono19}. }
\end{table*}

The population ratio (N$_c$/N$_{tot}$) is a solid parameter to trace the
topology of the instability strip, i.e. the region of the instability strip in
which fundamental and first overtone RRLs attain a stable limit cycle. The
dependence of the population ratio on metallicity provides not only quantitative
clues on the topology of the instability strip, but also on the dependence of
the HB morphology on the metal content.

The accuracy of the population ratio relies on the completeness limits of both
RRab and RRc variables. The RRc variables are more affected by an observational
bias than RRab variables. The RRLs obey to well defined Period-Luminosity
relations for wavelengths longer than the \rr-band and the RRc variables are
typically fainter than RRab variables. Moreover, the RRc luminosity amplitudes
are on average smaller than the RRab variables. To investigate on a more
quantitative basis this issue, Fig.~\ref{fig:Gratio} shows the comparison
between the apparent magnitude distribution of RRc variables (red histogram) and
the total number of RRLs. The two histograms display the same trend when moving
from the bright to the faint limit of the distribution. Thus suggesting that the
completeness of both RRab and RRc variables is similar over the entire magnitude
range.

The top panel of Fig.~\ref{fig:ratio} shows the population ratio (solid black
line) as a function of the iron content for Galactic RRLs, 
by using the running average algorithm described in Sect.~\ref{sec:metalbins},
together with its standard deviation (grey hatched area). Data plotted in this
figure bring forward several new features worth being discussed in detail.

{\em Continuous variation} -- The variation of the population ratio over the
entire metallicity range is continuous and does not show evidence of a
dichotomic distribution. Indeed, the population ratio steadily increases in the
metal-poor regime and it approaches a well defined plateau with
N$_c$/N$_{tot}$$\sim$0.36 for $-2.15\le\feh\le-1.70$. The trend is opposite in
the metal-intermediate regime, with the population ratio steadily decreasing and
it attains its absolute minimum (N$_c$/N$_{tot}$$\sim$0.18) for
$-0.9\le\feh\le-0.8$. Note that the standard deviation in this metallicity
regime is systematically smaller than in the metal-poor/metal-rich regime
because the mean of the metallicity distributions of both RRab and RRc (see the
bottom panel of the same figure) is located at $\feh\simeq-1.48$ and $\simeq
-1.58$, respectively. The population ratio shows, once again, a steady increase
in the metal-rich regime and it attains values close to its mean value
(N$_c$/N$_{tot}$$\sim$0.29) at solar iron abundance. Note that the increase in
the population ratio for iron abundances larger than $-0.8$~dex cannot be
explained by plain evolutionary arguments. Indeed, an increase in the metal
content causes a systematic drift of the HB morphology towards redder colours.
This means that the RRL instability strip is going to be mainly populated in its
redder region where only RRab variables attain a stable limit cycle. The
consequence is that the population ratio in the metal-rich regime should display
either a steady decrease or approaching an almost constant value. To further
constrain the behaviour of the population ratio as function of metallicity, we
also provide a continuous analytical fit (blue dashed line) by combining a
linear plus a sinusoidal function:
\begin{eqnarray}
\frac{N_c}{N_{tot}}&=& 0.212(\pm0.004)-0.047(\pm0.002)\cdot \feh+\nonumber\\
&& +0.073(\pm0.003)\cdot \sin\left(2.19(\pm0.08)\right.+\nonumber\\
&&\left.+3.78(\pm0.04)\cdot \feh\right) \;\;\;\; \left[\sigma=0.0002\right]
\label{eqn:sinfit}	
\end{eqnarray}

The variation of the population ratio described above could be affected by an
observation bias: the number of RRL variables increases when approaching the
Galactic plane. However, these regions are also severely affected by high
extinction. This means that the current RRL sample is far from being complete at
low Galactic latitudes \citep{preston91}. Moreover, the fraction of metal-rich
RRLs increases in approaching the Galactic plane \citep{layden94}. This means
that the continuous variation shown in Fig.~\ref{fig:ratio} is not affected by
this bias, but the accuracy concerning the metal-rich sub-sample will improve 
once more complete RRL sample will become available.

{\em Global trend} -- According to the Oosterhoff dichotomy we would expect to
have a population ratio of $\approx$0.44 for more metal-poor stellar systems
(OoII) and a value close to $\approx$0.3 for more metal-rich stellar systems
(OoI). However, data plotted in the top panel of Fig.~\ref{fig:ratio} are far
from being representative of the quoted dichotomic trend. The population ratio
attains, within the errors, relative maxima (N$_c$/N$_{tot}$$\sim$0.36), both in
the metal-poor and in the metal-rich regime.

{\em Comparison with globular clusters} -- To further constrain possible
similarities between field and cluster RRLs we compared the current population
ratio with similar estimates for Galactic globulars hosting more than 33 RRLs
\citep{clement01}. We arbitrarily selected this number of cluster RRLs to limit
statistical fluctuations in the population ratio. The population ratios of the
globulars are plotted in Fig.~\ref{fig:ratio} with different symbols (see
labelled names and values listed in Tab.~\ref{tbl:ggc}) and their trend is far
from being homogeneous. Cluster RRLs in the metal-poor regime ($\feh\le-1.6$)
either agree within the errors or display population ratios typical of OoII
stellar systems (marked with blue open circles) that are systematically larger
than field RRLs. 
The comparison between cluster and field RRLs is more complex in the
metal-intermediate regime ($-1.5\le\feh\le-1.0$), because the population ratio
for typical OoI stellar systems (marked with red open circles) ranges from less
than 0.1 for NGC~3201 to more than 0.5 for M14. In this metallicity regime the
field RRLs display a steady decrease from $\sim$0.3 to $\sim$0.2 and the
smallest dispersion.  
The comparison in the metal-rich ($\feh\ge-1$) regime is hampered by the limited
number of Galactic globulars hosting sizeable RRL samples. One out of the three
globulars present in this metallicity regime (NGC~6441) agrees quite well with
field RRLs, but NGC~6388 (with a similar iron abundance), and in particular,
NGC~6362 ($\sim$0.3~dex more metal-poor) attain population ratios that are
systematically larger than field ones. Note that the two metal-rich globulars
have HB morphologies dominated by an extended blue tail and by a sizeable sample
of red HB stars. In this context it is worth mentioning $\omega$~Cen, since it
hosts a sizeable sample of RRLs (191) and they display a well defined spread in
iron abundance ($-2.58\le\feh\le-0.85$, \citealt{magurno19}) with a population
ratio of 0.53. $\omega$~Cen is the most massive globular and its HB morphology
is dominated by an extended blue tail.     

To investigate on a more quantitative basis the possible correlation between the
population ratio and the HB morphology, the top panel of Fig.~\ref{fig:ratio3}
shows the population ratio as a function of the HB morphology index
[$HBR$=$(B-R)/(B+V+R)$] introduced more than 35 years ago by \citet[][see also
\citealt{lee94}]{lee89}. In the $HBR$ index the parameter $B$, $R$ and $V$ take
accounts for the number of HB stars that are either hotter ($B$) or cooler ($R$)
than the RRL instability strip, while $V$ is the number of RRLs. As expected
OoI/OoII stellar systems attain population ratios that are on average
smaller/larger than the mean Halo value (dotted line). The current data do not
display a global trend when moving from globulars with HB morphologies dominated
by red HB stars (on average more metal-rich and with negative $HBR$ values) and
globulars with HB morphologies dominated by blue HB stars (on average more
metal-poor and with positive $HBR$ values). There is evidence that the bulk of
OoI clusters distribute along a redder/bluer sequence departing from
$HBR\simeq-0.15$ and $HBR\simeq0.15$, respectively. The two sequences show
modest variations in the $HBR$ parameter, but the population ratio changes by
more than a factor of four. Indeed, two globulars of the redder sequence, namely
NGC~3201 and IC~4499 have similar $HBR$ values ($HBR$=$-0.04$ \textit{vs}
$-0.10$), but the former one only host a few RRc variables (7 out of 84,
N$_c$/N$_{tot}$=0.08) variables while the latter includes 35 RRc variables out
of 98 RRLs (N$_c$/N$_{tot}$=0.36). The same outcome applies to the bluer
sequence: the two globulars NGC~6934 and M62 have $HBR$ values of 0.13 and 0.32,
while the population ratio changes from 0.12 to 0.36. A similar sequence is also
present among OoII clusters, indeed, the two globulars M2 and M53 have similar
$HBR$ values ($HBR$=0.92 \textit{vs} 0.89), but the former one only host a few
RRc variables (15 out of 38, N$_c$/N$_{tot}$=0.39) while the latter includes
more RRc than RRab variables (35 out of 63, N$_c$/N$_{tot}$=0.56).

The morphology index $HBR$ presents several pros and cons when compared with the
morphology index $\tau_{\rm HB}$ recently introduced by \citet{torelli19}. This
new $\tau_{\rm HB}$ parameter is defined as the ratio between the areas
subtended by the cumulative number distribution in magnitude and in colour of
all the stars distributed along the HB. The bottom panel of
Fig.~\ref{fig:ratio3} shows the population ratio as a function of $\tau_{\rm
HB}$ for a sub-sample of the globulars adopted in the top panel. The current
data indicate a mild preliminary evidence of a steady increase in the population
ratio when moving from globulars with intermediate HB morphologies (the
horizontal portion of the HB, i.e., blue HB, red HB and the instability strip
are well populated) to globulars with HB morphologies dominated by blue stars.
The spread among the OoI clusters is still quite large, and indeed, the two
peculiar clusters with HB morphologies dominated by red HB stars (smaller
$\tau_{\rm HB}$ values), but with extended blue tails, attain larger population
ratios. This sub-group also includes four metal-intermediate clusters (M4,
IC~4499, NGC~6362, NGC~6584) with intermediate HB morphologies. The degeneracy
of these two sub-groups in the population ratio \textit{vs} $\tau_{\rm HB}$
plane needs to be investigated in more detail.    

It bears mentioning that the number of OoII globulars for which the $\tau_{\rm
HB}$ is available is still too limited to investigate in detail their
properties. Finally, we note that the two globulars with HB morphologies
dominated by blue HB stars include a metal-poor (M15, $\feh=-2.37$) and a metal-
intermediate (M2, $\feh=-1.65$) cluster with population ratios that agree quite
well with field RRLs with similar iron abundances.

The current circumstantial evidence further support that globulars played a
subdominant role in the build-up of the stellar Halo, also in agreement with the
results of recent observational surveys \citep{hanke20} and Galactic simulations
\citep{reinacampos20}.

\section{Summary and final remarks} 
\label{sec:summary}
We performed a new and homogeneous spectroscopic analysis of field Halo RRLs,
with iron abundances estimated using both low- and high-resolution spectra
collected at random phases. The iron abundances based on high-resolution spectra
include 190 RRLs \citet{crestani21b}, while those based on low-resolution
spectra include more than 7,700 RRLs. The latter sample is based on the new
calibration of the \deltaS\ method recently provided by \citet{crestani21}.
These abundances were complemented with iron abundances based on both low- and
high-resolution spectra available in the literature and brought to our
metallicity scale. To provide a homogeneous metallicity scale, special attention
was paid in the inclusion of the different datasets. The transformations into
our metallicity scale were estimated by using the objects in common between our
catalog and the datasets available in the literature.

We ended up with the largest spectroscopic catalog ever collected for both
fundamental and first overtone RRLs. The current sample includes 9,015 RRLs
(6,150 RRab, 2,865 RRc) with at least one metallicity estimate. Moreover and
even more importantly, the current spectroscopic catalog covers the extent of
the Galactic stellar Halo, with Galactocentric distances ranging from $\approx$5
to more than 140~kpc. This spectroscopic catalog was used to address several
pending issues concerning the pulsation properties of field RRLs and their use
as tracers of old stellar populations:

{\em Metallicity distribution function} -- The cumulative metallicity
distribution function (MDF) shows a mean value at
$\langle\feh\rangle=-1.51\pm0.01$ with a standard deviation $\sigma$=0.41~dex.
The current estimates agree with similar estimates available in the literature
and bring forward a clear asymmetry. Indeed, the MDF shows a long tail in the
metal-poor regime approaching $\feh\sim-3$ and a sharp metal-rich tail
approaching solar iron abundance. The large sample of spectroscopic measurements
allows us to investigate in detail the MDF of both RRab and RRc variables. We
found that RRab variables are systematically more metal-rich
($\langle\feh\rangle_{ab}=-1.48\pm 0.01$, $\sigma=0.41$~dex) than RRc variables
($\langle\feh\rangle_{c}=-1.58\pm 0.01$, $\sigma=0.40$~dex). This finding fully
supports preliminary estimates by \citet{liu20} and by \citet{crestani21} using
smaller datasets.   

A preliminary qualitative explanation of the difference in the MDF of RRab and
RRc variables can be provided by using plain evolutionary and pulsation
arguments. The topology of the instability strip and the dependence of the HB
morphology on the metal content indicate that RRc variables can be more easily
produced in the metal-poor than in the metal-rich regime. However, it is worth
mentioning that the MDF of RRc variables displays a metal-rich tail that is more
significant than the metal-rich tail of RRab variables.

{\em Bailey Diagram} -- The distribution of field RRLs in the Bailey diagram
(visual amplitude \textit{vs} logarithmic period) shows several interesting
features. An increase in the metal content causes a smooth and systematic shift
towards shorter periods for both RRab and RRc variables. The analysis of the
iso-contour across the Bailey diagram indicates that the relative fraction of
RRab variables located along the short-period sequence (more metal-rich) and the
long-period sequence (more metal-poor) is 80\%\ and 20\%. Interestingly enough,
the relative fractions for RRc variables have an opposite trend, namely 30\%\
(short-period) and 70\%\ (long-period), respectively.

{\em Dependence of pulsation periods and visual amplitudes on metallicity} --
The large sample of spectroscopic measurements allowed us to investigate on a
quantitative basis, for the first time, the dependence of pulsation periods and
visual amplitudes on metallicity. We found that the pulsation period of both
RRab and RRc variables shows a steady decrease when moving from the metal-poor
to the metal-rich regime. 
We derived new analytical relations and we found that an increase of 1~dex in iron 
content causes on average a decrease of $\approx$0.04~dex in the logarithmic period.
The trend concerning the visual amplitudes is similar. The analytical relations 
indicate that the visual amplitude of RRab variables is almost a factor of two 
less sensitive to the metal content than for RRc variables, indeed, the coefficient 
of the metallicity term decreases from --0.038 to --0.058.
The difference might be associated to a stronger impact of
non-linear phenomena on RRab than on RRc luminosity amplitudes. In spite of this
difference visual amplitudes and periods for both RRab and RRc variables do show
smooth distributions over the entire metallicity range. This evidence fully
supports the preliminary results concerning the nature of the Oosterhoff
dichotomy obtained by \citet{fabrizio19} using a smaller dataset of only RRab
variables. Indeed, they showed that the Oosterhoff dichotomy is an a natural
consequence of the lack of metal-intermediate Galactic globular clusters hosting
a sizeable sample of RRL stars.

{\em Impact of the metallicity on the Blazhko effect} -- The large and
homogeneous spectroscopic dataset allowed us to investigate the impact that the
iron abundance has on the occurrence of the Blazhko phenomenon. We found that
candidate Blazhko RRLs pulsating in the fundamental mode (177 RRab) appear to be
distributed across the OoI sequence, i.e. they seem to be either
metal-intermediate or metal-rich objects. The candidate Blazhko RRLs pulsating
in the first overtone (25 RRc) seem to show an opposite trend, because they seem
to be mainly located across the OoII sequence. The fraction of candidate Blazhko
RRab variables more metal-rich than $\feh=-1.5$ is 28\% while for RRc variables
is 9\%. However, the number of RRc variables with spectroscopic iron abundances
is still too limited to reach a ﬁrm conclusion concerning their dependence on
the metallicity.

{\em Dependence of the population ratio (N$_c$/N$_{tot}$) on metallicity} -- We
investigated the dependence of the population ratio on the metal content and we
found that the trend of field RRLs is more complex than expected on the basis of
similar estimates for globular clusters available in the literature. We found
that the population ratio steadily increases from $\sim$0.25 to $\sim$0.36 when
moving from the very metal-poor regime to $\feh\simeq-1.8$. Moreover, the
population ratio shows a decrease by a factor of two (0.36 \textit{vs} 0.18) and
a smaller dispersion, at fixed iron content, in the metal-intermediate regime
($-1.8\le\feh\le-0.9$). Finally, it shows once again a steady increase when
moving into the metal-rich regime, approaching a value of $\sim$0.3 at solar
iron abundance. The current findings appear to be at odds with pulsation and
evolutionary predictions, because the number of RRc variables should steadily
decreases when moving from the metal-poor/metal-intermediate regime into the
metal-rich regime.

Concerning cluster RRLs, we also investigated the occurrence of a possible
correlation between the population ratio and the HB morphology. We adopted two
different HB morphology indices ($HBR$, $\tau_{\rm HB}$) and selected clusters
hosting roughly three dozen of RRLs. We found that globulars distribute along
several sequences in which the $HBR$ index shows either minimal or modest
variations, but the population ratio changes by more than a factor of two/four.
These sequences appear both in OoI and in OoII clusters. 
Furthermore, the $\tau_{\rm HB}$ morphology index shows a mild correlation with
the population ratio when moving from globulars characterised by intermediate HB
morphologies typical of OoI clusters to globulars dominated by extended blue HB
tails, typical of OoII clusters. Once again, OoI clusters with similar
$\tau_{\rm HB}$ values display variations in the population ratio by more than a
factor of two.

The analysis of field and cluster RRLs provides two interesting findings:
\textit{a)} the population ratio is a promising diagnostic to further
investigate the fine structure of the HB even for clusters with similar HB
morphology indices; \textit{b)} the different trends between field and cluster
RRLs in the population ratio \textit{vs} metallicity plane indicate that
globular clusters played a minor role, if any, in building up the Halo.
The early formation and evolution of the Halo and the role played by 
nearby stellar systems will be addressed by using pulsation properties, 
kinematics and metallicity distributions of field RRLs in a forthcoming 
paper.

A famous motto suggests that novel approaches to attack, and possibly explain,
longstanding astrophysical problems open up more problems than they are able to
solve. The results of this investigation concerning the Oosterhoff dichotomy and
the population ratio moves along this path.  

\acknowledgements
\small
It is a real pleasure to thank the anonymous referee for her/his pertinent 
suggestions that improved the content and the readability of the paper.\\
This work has made use of data from the European Space Agency (ESA) mission {\it
Gaia} (\url{https://www.cosmos.esa.int/gaia}), processed by the {\it Gaia} Data
Processing and Analysis Consortium (DPAC,
\url{https://www.cosmos.esa.int/web/gaia/dpac/consortium}). Funding for the DPAC
has been provided by national institutions, in particular the institutions
participating in the {\it Gaia} Multilateral Agreement. This research has made
use of the GaiaPortal catalogues access tool, ASI - Space Science Data Center,
Rome, Italy (\url{http://gaiaportal.ssdc.asi.it}).\\ 
Guoshoujing Telescope (the Large Sky Area Multi-Object Fiber Spectroscopic
Telescope LAMOST) is a National Major Scientific Project built by the Chinese
Academy of Sciences. Funding for the project has been provided by the National
Development and Reform Commission. LAMOST is operated and managed by the
National Astronomical Observatories, Chinese Academy of Sciences.
Funding for the Sloan Digital Sky Survey (SDSS) has been provided by the Alfred
P. Sloan Foundation, the Participating Institutions, the National Aeronautics
and Space Administration, the National Science Foundation, the U.S. Department
of Energy, the Japanese Monbukagakusho, and the Max Planck Society. The SDSS Web
site is http://www.sdss.org/. The SDSS is managed by the Astrophysical Research
Consortium (ARC) for the Participating Institutions. The Participating
Institutions are The University of Chicago, Fermilab, the Institute for Advanced
Study, the Japan Participation Group, The Johns Hopkins University, Los Alamos
National Laboratory, the Max-Planck-Institute for Astronomy (MPIA), the
Max-Planck-Institute for Astrophysics (MPA), New Mexico State University,
University of Pittsburgh, Princeton University, the United States Naval
Observatory, and the University of Washington.\\
We would also like to acknowledge the financial support of INAF (Istituto
Nazionale di Astrofisica), Osservatorio Astronomico di Roma, ASI (Agenzia
Spaziale Italiana) under contract to INAF: ASI 2014-049-R.0 dedicated to SSDC.\\
M.M. and J.P.M. were partially supported by the National Science Foundation 
under grant No. AST1714534.\\
E.G., Z.P., H.L., B.L. acknowledge support by the Deutsche Forschungsgemeinschaft 
(DFG, German Research Foundation) – Project-ID 138713538 – SFB 881 
(“The Milky Way System”, sub-projects A03, A05 and A11).

\bibliographystyle{aasjournal}
\bibliography{biblio}


\end{document}

%% file: defs.tex
\newcommand{\feh}{\hbox{[Fe/H]}}

\newcommand{\hbeta}{\hbox{H$\beta$}}
\newcommand{\hgamma}{\hbox{H$\gamma$}}
\newcommand{\hdelta}{\hbox{H$\delta$}}
\newcommand{\caiik}{\hbox{Ca\,{\scriptsize II}\,K}}

\newcommand{\vv}{\hbox{\it V\/}}
\newcommand{\rr}{\hbox{\it R\/}}

\newcommand{\G}{\hbox{\it G\/}}

\newcommand{\deltaS}{\hbox{$\Delta$S}}

%% file: authors.tex
\author{V.F.~Braga}
\affil{INAF - Osservatorio Astronomico di Roma, Via Frascati 33, 00078, Monte Porzio Catone (Roma), Italy}
\affil{Space Science Data Center - ASI, Via del Politecnico s.n.c., 00133 Roma, Italy}

\author{J.~Crestani}
\affiliation{Dipartimento di Fisica, Universit\`a di Roma Tor Vergata, via della Ricerca Scientifica 1, 00133 Roma, Italy}
\affiliation{INAF -- Osservatorio Astronomico di Roma, via Frascati 33, 00078 Monte Porzio Catone, Italy}
\affiliation{Departamento de Astronomia, Universidade Federal do Rio Grande do Sul, Av. Bento Gon\c{c}alves 6500, Porto Alegre 91501-970, Brazil}

\author{G.~Bono}
\affiliation{Dipartimento di Fisica, Universit\`a di Roma Tor Vergata, via della Ricerca Scientifica 1, 00133 Roma, Italy}
\affil{INAF - Osservatorio Astronomico di Roma, Via Frascati 33, 00078, Monte Porzio Catone (Roma), Italy}


\author{I.~Ferraro}
\affil{INAF - Osservatorio Astronomico di Roma, Via Frascati 33, 00078, Monte Porzio Catone (Roma), Italy}

\author{G.~Fiorentino}
\affil{INAF - Osservatorio Astronomico di Roma, Via Frascati 33, 00078, Monte Porzio Catone (Roma), Italy}

\author{G.~Iannicola}
\affil{INAF - Osservatorio Astronomico di Roma, Via Frascati 33, 00078, Monte Porzio Catone (Roma), Italy}

\author{G.W.~Preston}
\affil{Carnegie Observatories, 813 Santa Barbara Street, Pasadena, CA 91101, USA}

\author{C.~Sneden}
\affil{Department of Astronomy and McDonald Observatory, The University of Texas, Austin, TX 78712, USA}

\author{F.~Th\'evenin}
\affil{Universit\'{e} de Nice Sophia-antipolis, CNRS, Observatoire de la C\^{o}te d'Azur, Laboratoire Lagrange, BP 4229, F-06304 Nice, France}

\author{G.~Altavilla}
\affil{INAF - Osservatorio Astronomico di Roma, Via Frascati 33, 00078, Monte Porzio Catone (Roma), Italy}
\affil{Space Science Data Center - ASI, Via del Politecnico s.n.c., 00133 Roma, Italy}

\author{B.~Chaboyer}
\affil{Department of Physics and Astronomy, Dartmouth College, Hanover, NH 03755, USA}

\author{M.~Dall'Ora}
\affil{INAF-Osservatorio Astronomico di Capodimonte, Salita Moiariello 16, 80131 Napoli, Italy}

\author{R.~da Silva}
\affil{INAF - Osservatorio Astronomico di Roma, Via Frascati 33, 00078, Monte Porzio Catone (Roma), Italy}
\affil{Space Science Data Center - ASI, Via del Politecnico s.n.c., 00133 Roma, Italy}

\author{E.~K. Grebel} 
\affiliation{Astronomisches Rechen-Instit\"ut, Zentrum f\"ur Astronomie der Universit\"at Heidelberg, M\"onchhofstr. 12-14, D-69120 Heidelberg, Germany}

\author{C.K.~Gilligan}
\affil{Department of Physics and Astronomy, Dartmouth College, Hanover, NH 03755, USA}

\author{H.~Lala}
\affiliation{Astronomisches Rechen-Instit\"ut, Zentrum f\"ur Astronomie der Universit\"at Heidelberg, M\"onchhofstr. 12-14, D-69120 Heidelberg, Germany}

\author{B.~Lemasle}
\affiliation{Astronomisches Rechen-Instit\"ut, Zentrum f\"ur Astronomie der Universit\"at Heidelberg, M\"onchhofstr. 12-14, D-69120 Heidelberg, Germany}

\author{D.~Magurno}
\affil{Dipartimento di Fisica e Astronomia "Augusto Righi", Universit\'a di Bologna, Viale Berti Pichat 6/2, 40127 Bologna, Italy}

\author{M.~Marengo}
\affil{Department of Physics and Astronomy, Iowa State University, Ames, IA 50011, USA}

\author{S.~Marinoni}
\affil{INAF - Osservatorio Astronomico di Roma, Via Frascati 33, 00078, Monte Porzio Catone (Roma), Italy}
\affil{Space Science Data Center - ASI, Via del Politecnico s.n.c., 00133 Roma, Italy}

\author{P.M.~Marrese}
\affil{INAF - Osservatorio Astronomico di Roma, Via Frascati 33, 00078, Monte Porzio Catone (Roma), Italy}
\affil{Space Science Data Center - ASI, Via del Politecnico s.n.c., 00133 Roma, Italy}

\author{C.~E.~Mart\'inez-V\'azquez}
\affiliation{Cerro Tololo Inter-American Observatory, NSF's National Optical-Infrared Astronomy Research Laboratory, Casilla 603, La Serena, Chile}

\author{N.~Matsunaga}
\affiliation{Department of Astronomy, The University of Tokyo, 7-3-1 Hongo, Bunkyo-ku, Tokyo 113-0033, Japan} 

\author{M.~Monelli}
\affil{Instituto de Astrof\'{i}sica de Canarias, Calle Via Lactea s/n, E-38200 La Laguna, Tenerife, Spain}
\affil{Departamento de Astrof\'{i}sica, Universidad de La Laguna, E-38200 La Laguna, Tenerife, Spain}

\author{J.~P. Mullen}
\affiliation{Department of Physics and Astronomy, Iowa State University, Ames, IA 50011, USA}

\author{J.~Neeley}
\affiliation{Department of Physics, Florida Atlantic University, 777 Glades Rd, Boca Raton, FL 33431 USA}

\author{M.~Nonino}
\affil{INAF-Osservatorio Astronomico di Trieste, Via G.B. Tiepolo, 11, I-34143 Trieste, Italy}

\author{Z.~Prudil}
\affiliation{Astronomisches Rechen-Instit\"ut, Zentrum f\"ur Astronomie der Universit\"at Heidelberg, M\"onchhofstr. 12-14, D-69120 Heidelberg, Germany}

\author{M~Salaris}
\affiliation{Astrophysics Research Institute, Liverpool John Moores University, IC2, Liverpool Science Park, 146 Brownlow Hill, Liverpool, L3 5RF, UK}

\author{P.~B.~Stetson}
\affiliation{Herzberg Astronomy and Astrophysics, National Research Council, 5071 West Saanich Road, Victoria, British Columbia V9E 2E7, Canada}

\author{E.~Valenti}
\affiliation{European Southern Observatory, Karl-Schwarzschild-Str. 2, 85748 Garching bei Munchen, Germany}
\affiliation{Excellence Cluster ORIGINS, Boltzmann-Stra\ss e 2, D-85748 Garching bei M\"{u}nchen, Germany}

\author{M.~Zoccali}
\affiliation{Instituto de Astrof\'sica, Facultad de F\'sica, Pontificia Universidad Cat\'lica de Chile, Av. Vicu\~na Mackenna 4860, Santiago, Chile}